\documentclass{aa}  
\usepackage{graphicx}
\usepackage{txfonts}
\usepackage{natbib}
\usepackage{color}
\usepackage{multirow}

\definecolor{red}{rgb}{1.,0.0,0.}

\newcommand{\logPOpPlmc}{1.19}
\newcommand{\logPOpPsmc}{1.52}
\newcommand{\logPOpPall}{1.18}
\newcommand{\logPOPR}{0.517}

\newcommand{\slopelmc}{-0.08}
\newcommand{\eslopelmc}{0.05}
\newcommand{\zerolmc}{1.22}
\newcommand{\ezerolmc}{0.02}
\newcommand{\cstelmc}{1.22}
\newcommand{\ecstelmc}{0.02}

\newcommand{\slopesmc}{0.07}
\newcommand{\eslopesmc}{0.12}
\newcommand{\zerosmc}{1.18}
\newcommand{\ezerosmc}{0.06}
\newcommand{\cstesmc}{1.20}
\newcommand{\ecstesmc}{0.05}

\newcommand{\slopeall}{-0.08}
\newcommand{\eslopeall}{0.04}
\newcommand{\zeroall}{1.24}
\newcommand{\ezeroall}{0.02}
\newcommand{\csteall}{1.24}
\newcommand{\ecsteall}{0.02}

\newcommand{\slopePR}{0.684}
\newcommand{\eslopePR}{0.007}
\newcommand{\zeroPR}{1.489}
\newcommand{\ezeroPR}{0.002}

\usepackage{natbib,twoopt}
\usepackage[breaklinks=true, colorlinks=true, citecolor = blue]{hyperref} %% to avoid \citeads line fills
\bibpunct{(}{)}{;}{a}{}{,} %% natbib format for A&A and ApJ
\makeatletter
\newcommandtwoopt{\citeads}[3][][]{\href{http://adsabs.harvard.edu/abs/#3}%
	{\def\hyper@linkstart##1##2{}%
		\let\hyper@linkend\@empty\citealp[#1][#2]{#3}}}
\newcommandtwoopt{\citepads}[3][][]{\href{http://adsabs.harvard.edu/abs/#3}%
	{\def\hyper@linkstart##1##2{}%
		\let\hyper@linkend\@empty\citep[#1][#2]{#3}}}
\newcommandtwoopt{\citetads}[3][][]{\href{http://adsabs.harvard.edu/abs/#3}%
	{\def\hyper@linkstart##1##2{}%
		\let\hyper@linkend\@empty\citet[#1][#2]{#3}}}
\newcommandtwoopt{\citeyearads}[3][][]%
{\href{http://adsabs.harvard.edu/abs/#3}
	{\def\hyper@linkstart##1##2{}%
		\let\hyper@linkend\@empty\citeyear[#1][#2]{#3}}}
\makeatother

\begin{document}

	\title{Observational calibration of the projection factor of Cepheids}
	\subtitle{IV. Period-projection factor relation of Galactic and Magellanic Cloud Cepheids}
	\titlerunning{The $p$-factor of LMC Cepheids}
	
	\author{ A.~Gallenne\inst{1},
		P.~Kervella\inst{2, 3},
		A.~M\'erand\inst{4},
		G.~Pietrzy\'nski\inst{5,6},
		W.~Gieren\inst{5,7},
		N.~Nardetto\inst{8}
		%					D.~Graczyk\inst{1,4},
		%					P.~Konorski\inst{3},
		%				 R.~I.~Anderson\inst{7,8}
		%				\and S.~Villanova\inst{1}
		\and B.~Trahin\inst{2, 3}
	}
	
	\authorrunning{A. Gallenne et al.}
	
	\institute{European Southern Observatory, Alonso de C\'ordova 3107, Casilla 19001, Santiago 19, Chile
		\and Unidad Mixta Internacional Franco-Chilena de Astronom\'ia, CNRS/INSU UMI 3386 and Departamento de Astronom\'ia, Universidad de Chile, Casilla 36-D, Santiago, Chile
		\and LESIA, Observatoire de Paris, PSL Research University, CNRS, Sorbonne Universit\'es, UPMC Univ. Paris 06, Univ. Paris Diderot, Sorbonne Paris Cit\'e, 5 Place Jules Janssen, 92195 Meudon, France
		\and European Southern Observatory, Karl-Schwarzschild-Str. 2, 85748 Garching, Germany
		\and Universidad de Concepci\'on, Departamento de Astronom\'ia, Casilla 160-C, Concepci\'on, Chile
		\and Nicolaus Copernicus Astronomical Centre, Polish Academy of Sciences, Bartycka 18, PL-00-716 Warszawa, Poland
		\and Millenium Institute of Astrophysics, Av. Vicu{\~n}a Mackenna 4860, Santiago, Chile
		\and Laboratoire Lagrange, UMR7293, Universit\'e de Nice Sophia-Antipolis, CNRS, Observatoire de la C\^ote dAzur, Nice, France
		%	%\and Observatoire de Gen\`eve, Universit\'e de Gen\`eve, 51 Ch. des Maillettes, 1290 Sauverny, Switzerland
		%	\and Department of Physics and Astronomy, Johns Hopkins University, Baltimore, MD 21218, USA
		%	\and D\'epartement d'Astronomie, Universit\'e de Gen\`eve, 51 Ch. des Maillettes, 1290 Sauverny, Switzerland
	}%\\
	
	\offprints{A. Gallenne} \mail{agallenne@astro-udec.cl}
	
	\date{Received July 17, 2017; accepted August 31, 2017}
	
	% \abstract{}{}{}{}{} 
	% 5 {} token are mandatory
	
	\abstract
	{The Baade-Wesselink (BW) method, which combines linear and angular diameter variations, is the most common method to determine the distances to pulsating stars. However, the projection factor, $p$-factor, used to convert radial velocities into pulsation velocities, is still poorly calibrated. This parameter is critical on the use of this technique, and often leads to 5-10\,\% uncertainties on the derived distances.}
	{We focus on empirically measuring the $p$-factor of a homogeneous sample of 29 LMC and 10 SMC Cepheids for which an accurate average distances were estimated from eclipsing binary systems.}
	{We used the SPIPS algorithm, which is an implementation of the BW technique. Unlike other conventional methods, SPIPS combines all observables, i.e. radial velocities, multi-band photometry and interferometry into a consistent physical modelling to estimate the parameters of the stars. The large number and their redundancy insure its robustness and improves the statistical precision.}
	{We successfully estimated the $p$-factor of several Magellanic Cloud Cepheids. Combined with our previous Galactic results, we find the following $P-p$ relation: $\slopeall_{\pm\eslopeall} (\log P - \logPOpPall) + \zeroall_{\pm\ezeroall}$. We find no evidence of a metallicity dependent $p$-factor. We also derive a new calibration of the period-radius relation, $\log R = \slopePR_{\pm\eslopePR} (\log P - \logPOPR) + \zeroPR_{\pm\ezeroPR}$, with an intrinsic dispersion of 0.020. We detect an infrared excess for all stars at $3.6\mu$m and $4.5\mu$m, which might be the signature of circumstellar dust. We measure a mean offset of $\Delta m_{3.6} = 0.057 \pm 0.006$\,mag and $\Delta m_{4.5} = 0.065 \pm 0.008$\,mag.}
	{We provide a new $P-p$ relation based on a multi-wavelength fit that can be used for the distance scale calibration from the BW method. The dispersion is due to the LMC and SMC width we took into account because individual Cepheids distances are unknown. The new $P-R$ relation has a small intrinsic dispersion: 4.5\,\% in radius. This precision will allow us to accurately apply the BW method to nearby galaxies. Finally, the infrared excesses we detect again raise the issue of using mid-IR wavelengths to derive period-luminosity relation and to calibrate the Hubble constant. These IR excesses might be the signature of circumstellar dust, and are never taken into account when applying the BW method at those wavelengths. Our measured offsets may give an average bias of $\sim2.8$\,\% on the distances derived through mid-IR $P-L$ relations.}
	
	\keywords{techniques: photometric, radial velocities --- stars: fundamental parameters --- stars: variables: Cepheids}
	\maketitle
	
	%________________________________________________________________

	\section{Introduction}
	
	Classical Cepheids are of fundamental importance for the extragalactic distance scale as they are the first rung of the cosmological distance ladder. The empirical relation between their pulsation period and intrinsic luminosity, called the period-luminosity relation \citep[P--L, or also Leavitt law,][]{Leavitt_1908__0, Leavitt_1912_03_0}, makes Cepheids very useful standard candles. A careful and accurate calibration of this relation is necessary as they are used to estimate distances to farther galaxies and derived cosmological parameters such as the Hubble constant, $H_0$ \citep[see e.g.][]{Riess_2011_04_0}.
	
	To calibrate the zero point of this relation, we need independent distance measurements of some Cepheids. The most common way for that is the Baade-Wesselink (BW) method, also called the parallax-of-pulsation method, which compares the angular diameter variations of the star (from surface brightness-colour relations or optical interferometry) with the linear diameter variations (from the integration of the radial velocity). The distance of the Cepheid is then obtained by fitting the linear and angular diameter amplitudes \citep[see e.g.][]{Gallenne_2012_03_0}. However, the distance is degenerate with the parameter called the projection factor, $p$-factor. The measured radial velocity, which needs to be integrated to obtain the linear diameter, is not the pulsation velocity, but its projection weighted by the intensity at each point of the stellar disk. Therefore, to convert the radial velocity into the pulsation velocity, we conventionally choose a multiplicative constant $p$-factor. This factor depends on the limb darkening and the dynamical behaviour of the line-forming regions, which is rather difficult to quantify without a detailed model of the stellar atmosphere. This parameter is currently the main source of uncertainty in the application of the BW method, leading to a global uncertainty of about 5-10\,\% on the distance.
	
	There is no agreement in the literature about the optimum value of the $p$-factor. Different authors use either a constant value (ranging from 1.2 to 1.5) or a linear dependence of the $p$-factor with the pulsation period ($P-p$ relation). However, that dependence differs between authors \citep[see e.g.][]{Gieren_2005_07_0,Nardetto_2007_07_0,Storm_2011_10_0,Neilson_2012_05_0,Nardetto_2014_01_0}, with a slope ranging from -0.05 to -0.19 and a zero point from 1.31 to 1.58, from theoretical or empirical relations (for a small sample of Galactic Cepheids). An observational calibration is possible if the distance to the Cepheid is known, as demonstrated in our previous works for a few Cepheids \citep{Merand_2005_04_0,Pilecki_2013_12_0,Breitfelder_2015_04_0,Gieren_2015_12_0}. Our more recent analyses with the Spectro-Photo-Interferometry of Pulsating Stars algorithm \citep[SPIPS,][]{Merand_2015_12_0} seem to show a mildly variable $p$-factor with respect to the pulsation period \citep{Breitfelder_2016_03_0,Kervella_2017_04_0}; the latest value is $p = 1.29 \pm 0.04$.
	
	In this paper, we focus on the observational calibration of the $p$-factor for Cepheids in the Magellanic Clouds, using the accurate  mean distance $d_\mathrm{LMC} = 49.97 \pm 1.13$\,kpc from \citet{Pietrzynski_2013_03_0}, and $d_\mathrm{SMC} = 62.1 \pm 1.9$\,kpc from \citet{Graczyk_2014_01_0}. The sample is larger than our previous Galactic work; here we use 29 LMC and 10 SMC Cepheids, which enables us to better constrain a possible $P-p$ relation. In Sect.~\ref{section__data_sample} we present the photometric and radial velocity data we retrieved from the literature; we perform a global fit in Sect.~\ref{section__spips_analysis}. We then discuss the results and make concluding remarks in Sects.~\ref{section__discussion} and \ref{section__conclusion}.

	%\begin{figure*}[!ht]
	%\centering
	%\resizebox{\hsize}{!}{\includegraphics{TZFor_RV.pdf}\includegraphics{TZFor_spec.pdf}
	%}
	%\caption{{\it Left}: The broadening function profiles of TZ For derived from HARPS spectrum taken on December 6, 2010. The primary rotates synchronously (high peak on left) and the secondary's profile is flattened because of fast rotation. {\it Right}: normalized, disentangled spectra of the primary (narrow lines) and the secondary (broad shallow lines) around of H$\beta$ line. }
	%\label{figure_rv}
	%\end{figure*}

	\section{Data sample}
	\label{section__data_sample}
	
	\begin{figure*}[!ht]
		\centering
		\resizebox{\hsize}{!}{\includegraphics{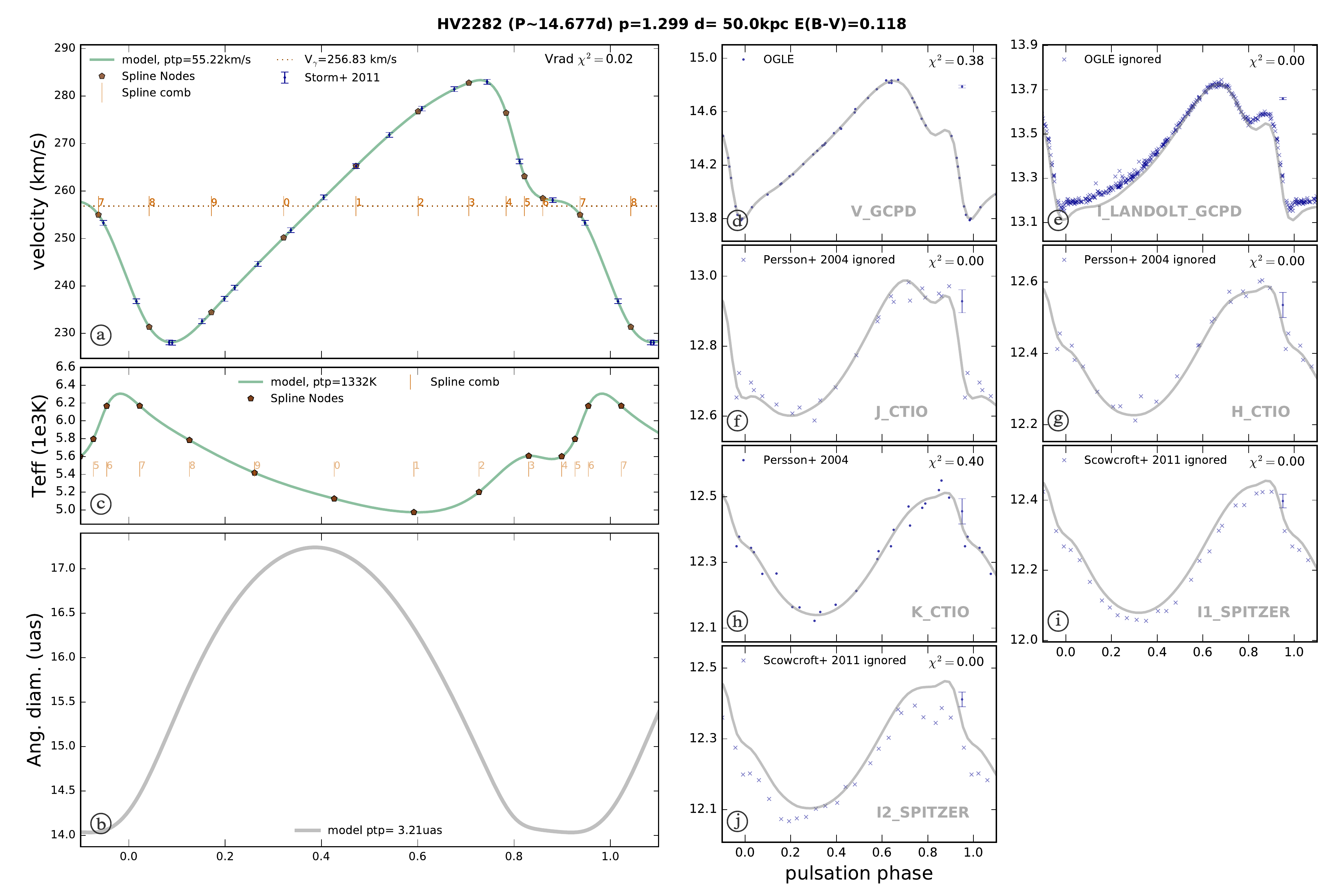}}
		\caption{Baade-Wesselink fit of the Cepheid HV2282 using only $V, K$ photometry with RVs to highlight the inconsistency of the OGLE $I$ photometry. The number $i$ in the spline comb denotes the $i$th node. Angular diameter in the bottom left panel is in $\mu$as. Y-axis for the photometry (right panels) is listed in magnitude. The points on the right side of the photometric plots give the error bar scale. Crosses mean that the points are ignored.}
		\label{figure_ogle_I_problem}
	\end{figure*}
	
	We selected Cepheids for which we have at least $V, J, H, K$ light curves and radial velocity measurements.  For the LMC, we used the sample of \citet{Storm_2011_10_1} for which precise velocities have been obtained with a good phase coverage. Their sample contains 22 LMC Cepheids; we added 7 LMC Cepheids having velocities reported by \citet{Imbert_1985_08_0,Imbert_1989_12_0}. The Cepheid HV900 is in common between the two datasets. For the SMC, we used 5 Cepheids from \citet{Storm_2004_02_1}, together with 5 additional stars from \citet{Imbert_1989_12_0}. We note that the SMC data have lower accuracy that the LMC data.
	
	$V$-band photometry was retrieved from the OGLE III database \citep{Soszynski_2008_09_0}. OGLE $I$-band measurements were not used because of an inconsistency between the models when using other photometric bands. From a multi-wavelength fit (see Sect.~\ref{section__spips_analysis}), for almost every star we found both a non-constant offset between the model and the measurements, and an incompatibility of the amplitude with respect to the pulsation phase. An example is shown in fig.~\ref{figure_ogle_I_problem}, for which we performed a 'classical' BW fit, i.e. using only $V, K$ photometry with RVs, while the other photometric data are ignored (but plotted). While the $J$ and $H$ bands match the model, this is not the case for the OGLE $I$ band. This might point to a problem with the characterization of this filter, such as an inaccurate zero point calibration or an incorrect effective wavelength. We note that we used the Landolt system, to which the OGLE photometry has been transformed. As pointed out by \citet{Udalski_2002_09_0}, one of the main sources of systematic errors could be the difference in transmission profile between the OGLE filters and the standard system, which can cause some non-linearities in transformations. Therefore, to avoid possible sources of bias, we decided not to use the $I$-band data. The use of the $I$ band would add a better leverage to the reddening determination, but incorrect photometric values would seriously bias its estimate, and would even induce infrared excess to longer wavelengths, mimicking the presence of a circumstellar envelope. We do not notice any systematic offset of the OGLE $V$ band with respect to the standard system.
	
	We gathered additional $B, V$ measurements from \citet{Madore_1975_06_0}, \citet{Eggen_1977_05_0}, \citet{Moffett_1998_07_0}, \cite{Martin_1979__0}, \citet{van-Genderen_1983_06_0} and \citet{Storm_2004_02_1} when available\footnote{Using the McMaster Cepheid Photometry and Radial Velocity Data Archive: \url{http://crocus.physics.mcmaster.ca/Cepheid/HomePage.html}}. $J, H$ and $K$ near-infrared photometric light curves from \citet{Persson_2004_11_0}, \citet{Laney_1986__0}, \citet{Welch_1984_04_0} and \citet{Storm_2004_02_1} were also collected. Finally, to complete the photometric sample, we also retrieved 3.5 and $4.5\mu$m light curves from Spitzer observations \citep{Scowcroft_2011_12_0}. It is worth mentioning that long-period Cepheids usually have better measurements as they are brighter, and therefore have higher signal-to-noise ratios.
	
	It is particularly effective to use $B, V$ observations to constrain the interstellar reddening, while infrared wavelengths give a stronger leverage for detecting circumstellar envelopes \citep[see i.e][for Galactic Cepheids]{Gallenne_2014_02_0,Gallenne_2013_10_0,Kervella_2013_02_0,Gallenne_2011_11_0}

	\section{SPIPS analysis}
	\label{section__spips_analysis}
	
	\begin{figure}[!ht]
	\centering
	\resizebox{\hsize}{!}{\includegraphics{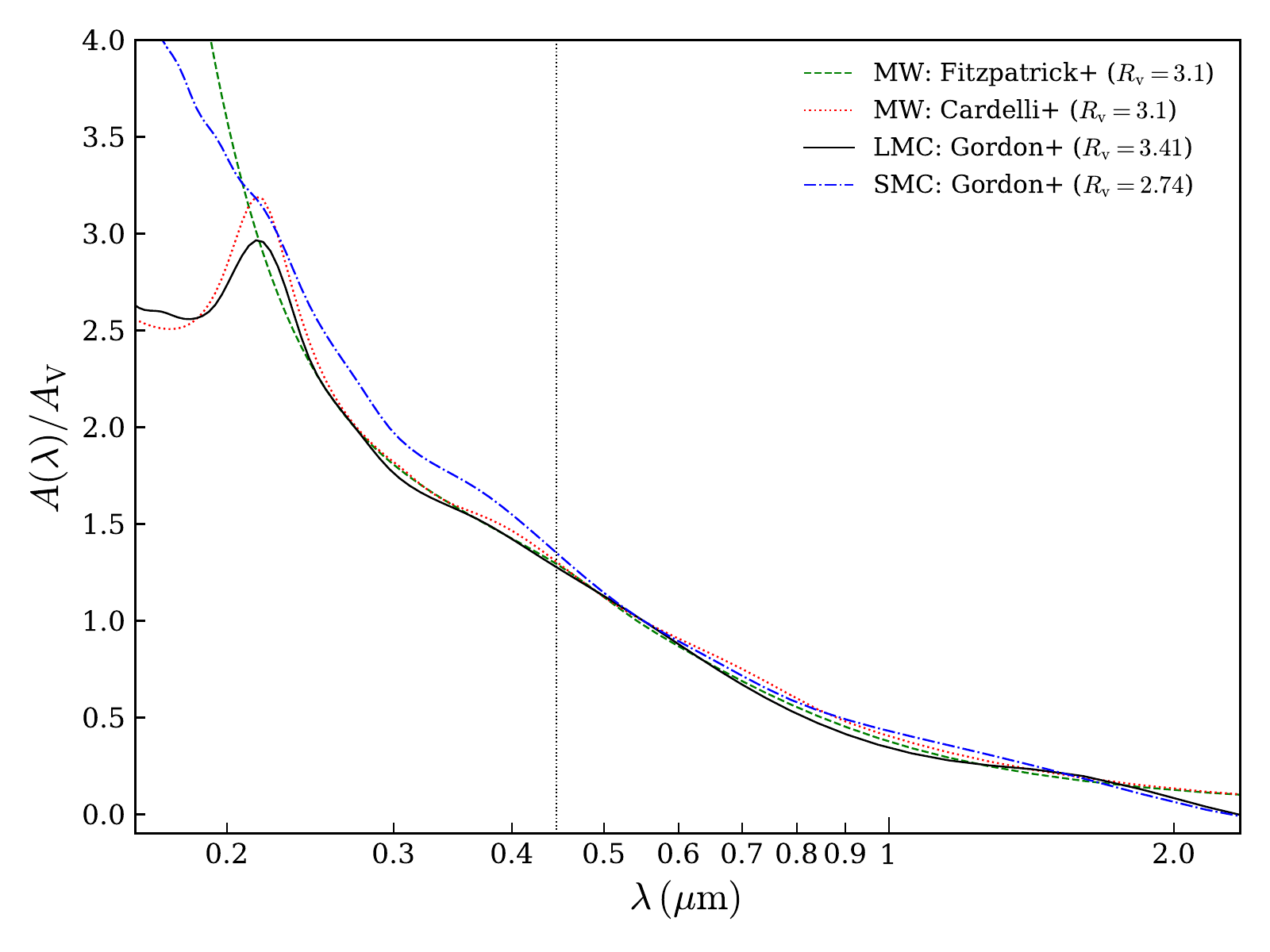}
	}
	\caption{Galactic \citep{Fitzpatrick_1999_01_0} and Large and Small Magellanic Cloud \citet{Gordon_2003_09_0} extinction curves.}
	\label{figure_reddening}
\end{figure}
	
	\subsection{Algorithm}

	We used the SPIPS modelling tool to perform a global fit of the radial velocities and multi-band photometry. A full description of the code can be found in \citet{Merand_2015_12_0}\footnote{The code is publicly available at \url{https://github.com/amerand/SPIPS}}. Briefly, this tool is based on the parallax-of-pulsation method (also called the Baade-Wesselink method), i.e. it compares the linear and angular variations of the Cepheid diameters to retrieve physical parameters of the stars, such as the ratio $p$-factor/distance, effective temperature, infrared excess, colour excess, etc. The SPIPS code can take several types of data and observables, e.g. optical and IR magnitudes and colours, radial velocities, and interferometric angular diameters. The resulting redundancy in the observables ensures a high level of robustness. We already proved the efficiency of SPIPS for type I and II Galactic Cepheids \citep{Kervella_2017_04_0,Breitfelder_2016_03_0,Breitfelder_2015_04_0}.
	
	Synthetic photometry is created using the ATLAS9 models \citep{Castelli_2003__0}, interpolated with Fourier series or periodic splines functions, and integrated using the bandpass and zero point of each observing filter\footnote{using the SVO and Asiago databases: \url{ http://svo2.cab.inta- csic.es/} and \url{http://ulisse.pd.astro.it/Astro/ADPS/Paper/index. html}.}. For LMC Cepheids, we interpolated the models for a metallicity of $-0.34$\,dex \citep{Luck_1998_02_0}.

	\subsection{Colour excess and circumstellar envelope}
	
	The reddening law and coefficient implemented in SPIPS are Galactic, taken from \citet{Fitzpatrick_1999_01_0} with $R_\mathrm{V} = 3.1$. There is evidence in the literature that the LMC and SMC reddening (law and coefficient) are different, more specifically for wavelengths shorter than the $B$ band. However, as shown in Fig.~\ref{figure_reddening}, the normalized extinction curves are qualitatively similar to our Milky Way (MW) for wavelengths longer that $0.4\,\mu$m. This is discussed in \citet{Gordon_2003_09_0}, who used a comparison of extinction curves to derive $R_\mathrm{v}$ = 3.41 for the LMC and $R_\mathrm{v}$ = 2.74 for the SMC. We therefore used these coefficients (not fitted) together with the reddening law from \citet{Fitzpatrick_1999_01_0}.
	
	We know that some Galactic Cepheids have some IR emission which can bias the photometric estimates \citep[see e.g.][]{Gallenne_2013_10_0,Gallenne_2011_11_0,Gallenne_2011_03_0,Barmby_2011_11_0}. The SPIPS code allows fitting the IR excess, either using a simple offset for each wavelength or  an analytical formula. However, due to our limited dataset which is not extremely accurate and the correlation with the colour excess, we assumed no IR excess in our SPIPS modelling. The fit of the stellar atmospheric models is performed from $B$ to $K$ band, but do not include the Spitzer measurements. Therefore, the presence of mid-IR excess will appear as an offset between the model and the data. This is discussed in Sect.~\ref{section__discussion}.
	
	\subsection{Distance to the LMC and SMC}
	The main issue in the parallax-of-pulsation method is the degeneracy between the distance $d$ and the p-factor $p$. \citet{Pietrzynski_2013_03_0} and \citet{Graczyk_2014_01_0} accurately derived the average distance of the Magellanic Clouds from eclipsing binary systems. We can therefore adopt this distance as a fixed parameter to derive the $p$-factor. However, the Cepheids can be located in front of or behind this average value. Individual distances are not possible to estimate. From three-dimensional maps using red-clump stars and Cepheids, the average depth of the LMC ranges from 3.4 to 4\,kpc, and from 4.2 to 4.9\,kpc for the SMC \citep{Subramanian_2009_03_0, Haschke_2012_10_1}. We adopted the value of 4\,kpc and 4.9\,kpc as a systematic error to the $p$-factor (added quadratically) respectively for the LMC and SMC.
	
	\subsection{Binary Cepheids}
	
	The LMC Cepheids HV914 and HV883 have been identified as spectroscopic binary because of variations in their systemic velocity. The SMC stars HV837 and HV11157 have also been identified as spectroscopic binary \citep{Szabados_2012_11_1,Imbert_1994_05_0}. The presence of a companion may bias the photometry of the Cepheid and offset the radial velocities. Therefore, stars with companions were not included in the final analysis.
	
	\subsection{$p$-factor determination}
	
	\begin{table*}[!h]
		\centering
		\caption{$p$-factor and reduced $\chi^2$ derived using fitted or fixed colour excess for the BW analysis.}
		\begin{tabular}{ccccc|cc}
			\hline\hline
			Star   & $\log P$  & \multicolumn{3}{c|}{fitted $E(B-V)$}	& \multicolumn{2}{c}{fixed $E(B-V)$} \\
					& 				& $p$\tablefootmark{a}   & $E(B-V)$  & $\chi^2_\mathrm{r}$ & $p$	& $\chi^2_\mathrm{r}$ \\
			\hline
			\multicolumn{7}{c}{LMC} \\
			\object{HV6093}  & 0.680 &  1.40$\pm$0.17 &  0.022$\pm$0.017 &  2.17  &  1.45$\pm$0.17 &  2.26  \\ 
			\object{HV2405}  & 0.840 &  1.39$\pm$0.17 &  0.051$\pm$0.012 &  1.11  &  1.38$\pm$0.17 &  1.15  \\ 
			\object{HV12452}  & 0.941 &  1.31$\pm$0.11 &  0.128$\pm$0.012 &  0.74  &  1.30$\pm$0.11 &  0.83  \\ 
			\object{HV12717}  & 0.947  &  1.34$\pm$0.12 &  0.090$\pm$0.013 &  1.10  &  1.37$\pm$0.12 &  1.11  \\ 
			\object{HV2527}  & 1.112 &  1.15$\pm$0.06 &  0.110$\pm$0.013 &  0.93  &  1.15$\pm$0.06 &  1.15  \\ 
			\object{HV2538}  & 1.142  &  1.37$\pm$0.08 &  0.120$\pm$0.012 &  0.79  &  1.38$\pm$0.09 &  0.87  \\ 
			\object{HV899}  & 1.492 &  1.18$\pm$0.05 &  0.037$\pm$0.012 &  2.88  &  1.16$\pm$0.05 &  3.00  \\ 
			\object{HV1006}  & 1.153 &  1.27$\pm$0.06 &  0.115$\pm$0.013 &  1.52  &  1.27$\pm$0.07 &  1.84  \\ 	
			\object{HV5655}  & 1.153 &  1.14$\pm$0.05 &  0.092$\pm$0.010 &  1.17  &  1.15$\pm$0.05 &  1.18  \\ 		
			\object{HV12505}  & 1.158  &  1.16$\pm$0.08 &  0.159$\pm$0.014 &  1.81  &  1.21$\pm$0.08 &  2.09  \\ 
			\object{HV2282}  & 1.167 &  1.31$\pm$0.05 &  0.118$\pm$0.009 &  0.51  &  1.32$\pm$0.05 &  0.55  \\ 
			\object{HV2549}  & 1.210  &  1.34$\pm$0.07 &  0.055$\pm$0.010 &  1.01  &  1.33$\pm$0.07 &  1.04  \\ 
			\object{HV1005}  & 1.272 &  1.25$\pm$0.05 &  0.091$\pm$0.013 &  1.29  &  1.25$\pm$0.05 &  1.30  \\ 
			\object{U1}  &  1.353 &  1.08$\pm$0.04 &  0.173$\pm$0.014 &  1.28  &  1.11$\pm$0.05 &  1.82  \\ 			
			\object{HV876}  & 1.356  &  1.22$\pm$0.05 &  0.080$\pm$0.038 &  1.62  &  1.21$\pm$0.05 &  1.60  \\ 
			\object{HV878}  & 1.367  &  1.19$\pm$0.05 &  0.053$\pm$0.013 &  1.74  &  1.18$\pm$0.05 &  1.78  \\ 
			\object{HV1023}  & 1.424 &  1.21$\pm$0.05 &  0.119$\pm$0.013 &  2.53  &  1.21$\pm$0.05 &  2.68  \\ 
			\object{HV873}  & 1.537  &  1.14$\pm$0.03 &  0.153$\pm$0.010 &  0.66  &  1.15$\pm$0.04 &  0.86  \\ 
			\object{HV881}  & 1.553  &  1.20$\pm$0.06 &  0.043$\pm$0.013 &  1.85  &  1.19$\pm$0.06 &  1.93  \\ 
			\object{HV879}  & 1.566  &  1.15$\pm$0.05 &  0.083$\pm$0.016 &  10.23  &  1.15$\pm$0.05 &  10.21  \\ 
			\object{HV909}  & 1.575  &  1.30$\pm$0.05 &  0.032$\pm$0.009 &  1.70  &  1.30$\pm$0.05 &  1.88  \\ 
			\object{HV2257}  & 1.596 &  1.14$\pm$0.05 &  0.070$\pm$0.012 &  3.57  &  1.14$\pm$0.05 &  3.57  \\ 
			\object{HV2338}  & 1.625  &  1.19$\pm$0.05 &  0.027$\pm$0.010 &  2.54  &  1.17$\pm$0.05 &  2.80  \\ 
			\object{HV877}  & 1.655  &  1.22$\pm$0.07 &  0.089$\pm$0.013 &  2.74  &  1.22$\pm$0.06 &  2.74  \\ 
			\object{HV900}  & 1.676  &  1.30$\pm$0.05 &  0.031$\pm$0.011 &  2.72  &  1.29$\pm$0.05 &  2.89  \\ 
			\object{HV2369}  & 1.685 &  1.17$\pm$0.05 &  0.102$\pm$0.013 &  2.71  &  1.19$\pm$0.05 &  2.73  \\ 
			\object{HV2827}  & 1.897 &  1.24$\pm$0.08 &  0.105$\pm$0.012 &  2.00  &  1.26$\pm$0.09 &  2.03  \\ 
			\hline
			\multicolumn{7}{c}{SMC}  \\
			\object{HV824}  & 1.130 &  1.29$\pm$0.07 &  0.041$\pm$0.018 &  1.30  &  1.31$\pm$0.07 &  1.30  \\
			\object{HV1335}  & 1.158  &  1.14$\pm$0.08 &  0.012$\pm$0.015 &  3.07  &  1.12$\pm$0.08 &  3.31  \\			
			\object{HV822}  & 1.200  &  1.20$\pm$0.09 &  0.008$\pm$0.012 &  6.97  &  1.14$\pm$0.09 &  7.43  \\
			\object{HV1328}  & 1.212  &  1.11$\pm$0.12 &  0.036$\pm$0.016 &  5.17  &  1.11$\pm$0.12 & 5.18  \\
			\object{HV1333}  & 1.224  &  1.13$\pm$0.05 &  0.048$\pm$0.012 &  1.83  &  1.13$\pm$0.05 &  1.83  \\
			\object{HV1345}  & 1.816  &  1.47$\pm$0.17 &  -0.008$\pm$0.018 &  2.84  &  1.46$\pm$0.18 &  2.90  \\
			\object{HV821}  & 1.932 &  1.20$\pm$0.06 &  0.032$\pm$0.018 &  2.11  &  1.19$\pm$0.06 &  2.13  \\ 
			\object{HV829}  & 2.105 &  1.24$\pm$0.08 &  0.092$\pm$0.018 &  1.57  &  1.27$\pm$0.08 &  1.59  \\
			\hline& 
		\end{tabular}
		\tablefoot{a) The quoted errors are only statistical, a 8\,\% error has to be added to take into account the average width of the LMC and SMC.}
		\label{table__chi2}
	\end{table*}
	
	For each Cepheid, we fitted the radial velocity and photometric curves using spline functions. Although this method is numerically less stable than Fourier series, it leads to smoother models and avoids the introduction of unphysical oscillations when the data do not have a good phase coverage. We made tests by fitting Fourier series to the radial velocities, and no variations larger than 0.5\,\% were found to the final derived parameters.
	
	We performed two distinct analyses. We first left $E(B-V)$ as a free parameter, together with the reference epoch of the maximum light $T_0$, the pulsation period $P$, and the rate of period change $\dot{P}$. We then kept the colour excess fixed to a mean value of 0.08 and 0.06 \citep{Caldwell_1985_02_0,Caldwell_1991__0} respectively for the LMC and SMC. The results are listed in Table~\ref{table__chi2}, together with our fitted $p$-factor. We see that there is a slight improvement in the $\chi^2$ when fitting the colour excess. As this is more consistent than allowing a different extinction for each star, we chose for the rest of the paper the results for which we allowed a fitted value. First guess values of $T_0$ and $P$ have been taken from \citet{Samus_2009_01_0}. The final SPIPS adjustments for two LMC and two SMC Cepheids are shown in Appendix~\ref{appendix__1}.

	\section{Discussion}
	\label{section__discussion}
	
	\subsection{Linear fit and uncertainties}
	\label{subsection__linear_fit_and_uncertainties}
	We compare different $P-p$ relations, a constant and a linear fit to our data, and we compare our results with others. For this reason, we place particular care in the uncertainty of the parameters we derive for our linear fit. The linear fit is parametrized as:
	\begin{displaymath}
		p = a\,(\log P - \log P_0) + b,
	\end{displaymath}
	where $a$ and $b$ are the slope and zero point respectively. One might argue that the choice of $\log P_0$ is arbitrary, but in the context of least-squares minimization, there are two advantages for choosing $\log P_0$ such that $\partial^2 \chi^2/\partial a \partial b$ is zero ($\chi^2$ being the distance between the model and the data). In this case, and only in this case, we have the two following properties. First, the extrapolated uncertainty of the linear model $p$, at any $\log P$, is simply expressed as
	\begin{equation}
		\sigma_p^2 = (\log P - \log P_0)^2\sigma_a^2 + \sigma_b^2.
		\label{eq_sigma2_p}
	\end{equation} 
	The general form of this equation includes an additional $2(\log P - \log P_0)\rho\sigma_a\sigma_b$ on the right hand side, where $\rho$ is the correlation between $a$ and $b$. Since  $\rho \propto \partial^2 \chi^2/\partial a \partial b$, this term is nullified by definition of $\log P_0$. Second, the uncertainty $\sigma_b$ on $b$ is minimized (whereas the uncertainty $\sigma_a$ on $a$ is independent on the choice of $\log P_0$). This is an important property to consider if one aims to compare different linear fits because not choosing the optimum $\log P_0$ will overestimate the uncertainty on the zero point.
	
	Considering our sample of $(p_i\pm e_i, log P_i)$  for $i$ in $1...N$, the $L_2$ distance we need to minimize is
	\begin{displaymath}
		\chi^2 = \sum_{i=1}^{i=N} \frac{\left[a\,(\log P_i - \log P_0) + b - p_i\right]^2}{e_i^2},
	\end{displaymath}
	and the condition $\partial^2 \chi^2/\partial a \partial b=0$ becomes
	\begin{equation}
		\log P_0 = \frac{\sum_{i=1}^{i=N} \log P_i/e_i^2} {\sum_{i=1}^{i=N} 1/e_i^2},
		\label{eq_logp0}
	\end{equation}
	which is a weighted average of the sample of $\log P_i$.
	
	The choice of $\log P_0$ ultimately depends on the dataset and is important in order to compare different linear fit in terms of prediction uncertainties. Every dataset has a different coverage in $\log P$ and the accuracy on the predicted $p$ depends on this coverage, as is captured in Eq.~\ref{eq_sigma2_p}, which is only valid for $\log P_0$ determined according to Eq.~\ref{eq_logp0}. The failure to publish $\log P_0$ alongside the values of $a\pm\sigma_a, b\pm\sigma_b$ prevents comparison of linear laws derived from datasets with different $\log P$ coverage. Alternatively, the correlation factor $\rho$ between $a$ and $b$ can be published so the complete form of Eq.~\ref{eq_sigma2_p} can be used:
	\begin{displaymath}
		\sigma_p^2 = \log P ^2\sigma_a^2 + \sigma_b^2 + 2\,\rho\,\sigma_a \sigma_b \log P
	\end{displaymath}
	For most (if not all) published works, this value of $\rho$ is missing. We can easily derive, for any dataset $(p_i\pm e_i, log P_i)$ and any choice of $\log P_0$, the analytical expression for $\rho$:
	\begin{equation}
		\rho = \frac{-\partial^2\chi^2/\partial a \partial b}{\sqrt{(\partial^2\chi^2/\partial a^2)(\partial^2\chi^2/\partial b^2)}} = -\frac{\sum_i \frac{\log P_i-\log P_0}{e_i^2}}{\sqrt{\sum_i \frac{(log P_i-\log P_0)^2}{e_i^2} \sum_i  \frac{1}{e_i^2}}}
		\label{eq_rho}
	\end{equation}
	
	We can estimate an order of magnitude, using simple assumptions: all error bars on the values of $p_i$ (the $e_i$'s) are equal, and $\log P_0=0$. Then, in that particular case, Eq.~\ref{eq_rho} simplifies to
	\begin{displaymath}
		\rho = -\frac{\sum_i \log P_i}{\sqrt{N\sum_i (\log P_i)^2 }}
	\end{displaymath}
	%To illustrate how ill advised it is to center the linear fit at $\log P_0=0$ for realistic datasets, we can compute $\rho=-0.94$ for a dataset of $\log P_i$ uniformly sampled between 0.5 and 2, which is close to our own sample and typical of other studies.
	
	\subsection{$P-p$ relation}
	
	Uncertainties in the $p$-factor is the main reason of the exclusion of BW-based distance of Galactic Cepheids from the Hubble constant determination \citep{Riess_2009_07_0}.  \citet{Benedict_2007_04_0} obtained a set of ten parallax measurements to Galactic Cepheids using the Fine Guidance Sensor of the Hubble Space Telescopes, however, the average precision is $\sim 8$\,\%. These measurements were used in our previous SPIPS analysis \citep{Kervella_2017_04_0,Breitfelder_2016_03_0} to determine their $p$-factor. The accuracy of the first data release of Gaia-TGAS \citep{Lindegren_2016_11_0} is too low to provide an accurate calibration.
	
	Our sample of 29 LMC and 10 SMC Cepheids is the largest so far allowing the empirical trend of the $p$-factor to be probed with respect to the pulsation period of the Cepheids. We start by analysing Cepheids separately, and then finally combine them. We note that all Cepheids presented here are classical Cepheids, and provide a uniform sample.
	
\begin{figure}[!ht]
		\centering
		\resizebox{\hsize}{!}{\includegraphics{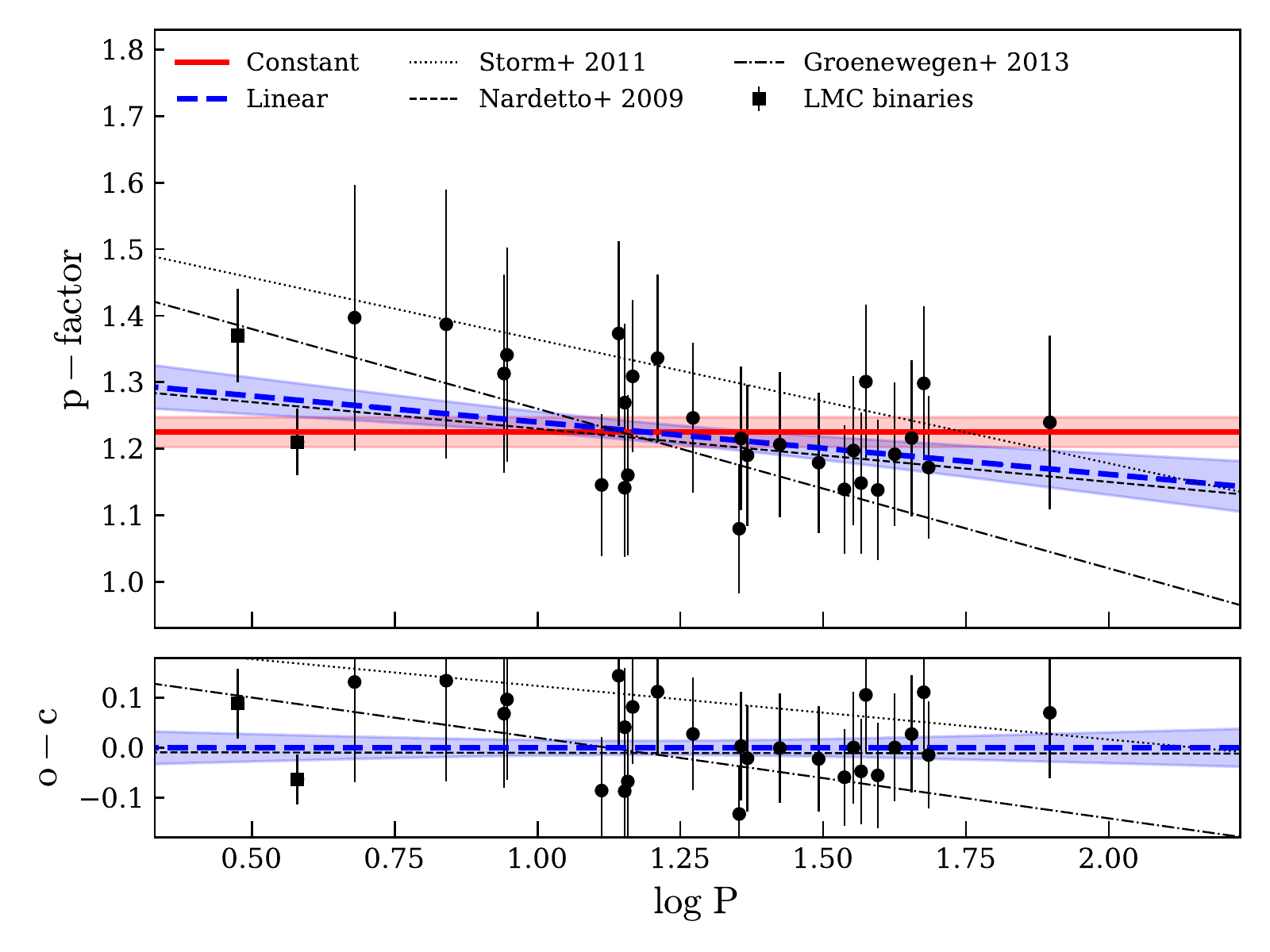}}
		\caption{Distribution of the LMC $p$-factor with respect to the pulsation period. The shaded blue and red areas denote the $1\sigma$ confidence interval, computed according to Eq.~\ref{eq_sigma2_p} in the case of the linear fit.}
		\label{figure_pfactor}
	\end{figure}

	\paragraph{- LMC only}: In Fig.~\ref{figure_pfactor}, we plotted the distribution of our estimated $p$-factor, whose error bars contain the $\sim 8$\,\% error from the average width. We also added two other $p$-factor values derived from two LMC Cepheids in an eclipsing binary system \citep[\object{OGLE LMC562.05.9009} and \object{OGLE LMC-CEP-0227},][]{Gieren_2015_12_0,Pilecki_2013_12_0}.
	
	We performed two fits: a linear law, as explained in Sect.~\ref{subsection__linear_fit_and_uncertainties}, and a constant $p$-factor. The result of the linear model gives $\slopelmc_{\pm\eslopelmc} (\log P - \logPOpPlmc) + \zerolmc_{\pm\ezerolmc}$ ($\sigma = 0.08$) with a $\chi_r^2 = 0.51$. The fit of a constant value gives $p = \cstelmc \pm \ecstelmc$ with $\chi_r^2 = 0.59$ (we took the standard error $\sqrt{ \sum_i \sigma_i^2 / N}$). A constant law does not seem to substantially change the fit quality, although the error bars are overestimated. To check whether a constant $p$ is statistically significant, we can perform an $F$-test and verify if the associated probability value is larger or smaller than 0.05. We found 0.03, meaning that the linear model is more statistically significant. The constant and linear models are plotted in Fig.~\ref{figure_pfactor}.
	
	We also plotted three other $P-p$ relations from the literature. From the measured atmospheric velocity gradient, \citet{Nardetto_2007_08_0,Nardetto_2009_05_0} derived a Galactic relation, $-0.08_{\pm0.05} \log P  + 1.31_{\pm0.06}$, which is nicely consistent with our LMC relation. This also seems to be consistent with the conclusion of \citet{Nardetto_2011_10_0} from hydrodynamical models, i.e. the $P-p$ relation of LMC Cepheids does not differ from the Galactic relation. The Galactic relation of \citet{Storm_2011_10_0}, $-0.186_{\pm0.06} \log P + 1.55_{\pm0.04}$, is in marginal agreement with Nardetto’s relation and ours, the slope being consistent within the combined uncertainties. Although they used strong constraints to determine this relation (i.e. the LMC distance is independent of the Cepheid pulsation period, and the HST distances should be equal to those from the BW technique), it is based on the IRSB method using only two photometric bands. \citet{Merand_2015_12_0} discussed an example using the SPIPS algorithm with several bands and a subset of photometric bands. They showed an agreement in the distance of $\eta$~Aql with Storm's value when using the same subset of data, but pointed out the underestimated statistical uncertainties when a subset is used. Another possible clue about this disagreement is the pulsation phase selection of \citet{Storm_2011_10_0}, i.e. they disregarded pulsation phases between 0.8 and 1 because of deviations between photometric and spectroscopic angular diameters. But these phases correspond to the minimum of the diameter, so ignoring them probably provides a biased estimate of the projection factor. \citet{Breitfelder_2016_03_0} apply the SPIPS method (with $B, V, J, H$, and $K$ photometric bands) to the same sample of nine Cepheids and instead found a constant Galactic $p$-factor, later confirmed by \citet{Kervella_2017_04_0} who found $p = 1.29 \pm 0.04$. However, their sample is smaller (only 11 stars). The relation of \citet{Groenewegen_2013_02_0} is also in disagreement, and is steeper by $\sim 3\sigma$. The discrepancy might also be explained by the use of only two photometric bands and an unfitted reddening. For instance, their assumed $E(B-V)$ values for $\delta$~Cep and $\ell$~Car are about two times higher than the fitted values of \citet[][$0.032\pm0.016$ for $\delta$~Cep]{Merand_2015_12_0} and  \citet[][$0.084\pm0.017$ for $\ell$~Car]{Breitfelder_2016_03_0}.

\begin{figure}[!ht]
	\centering
	\resizebox{\hsize}{!}{\includegraphics{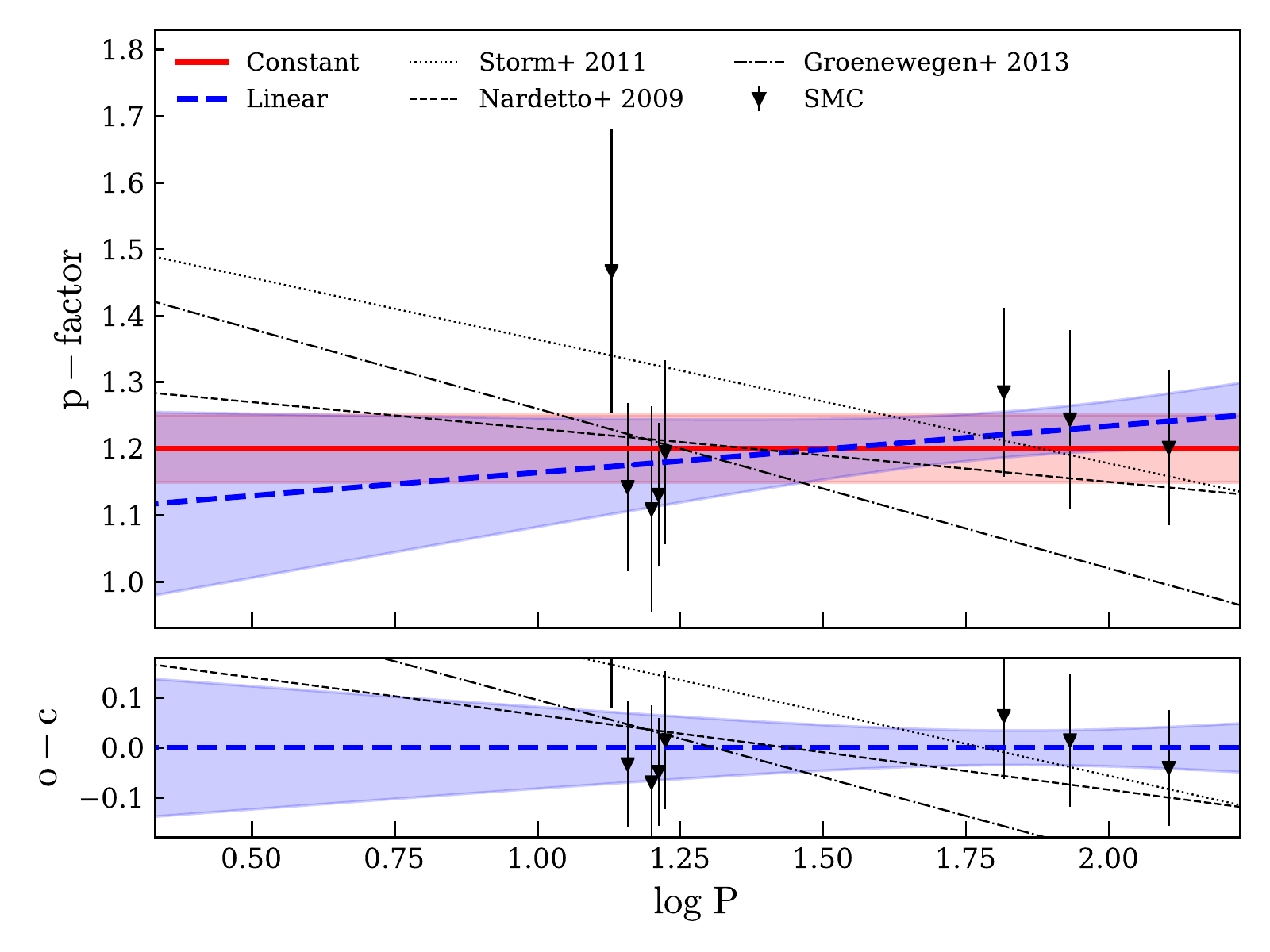}}
	\caption{Distribution of the SMC $p$-factor with respect to the pulsation period. The shaded blue and red areas denote the $1\sigma$ confidence interval, computed according to Eq.~\ref{eq_sigma2_p} in the case of the linear fit.}
	\label{figure_pfactor_smc}
\end{figure}

	\paragraph{- SMC only}: We performed the same analysis using the eight SMC Cepheids. The linear model gives $\slopesmc_{\pm\eslopesmc} (\log P - \logPOpPsmc) + \zerosmc_{\pm\ezerosmc}$ ($\sigma = 0.11$) with a $\chi_r^2 = 0.46$, while a constant fit provides $p = \cstesmc \pm \ecstesmc$ with $\chi_r^2 = 0.44$. The models are plotted in Fig.~\ref{figure_pfactor_smc}. We can see that the constant model is more consistent, in agreement with our estimated probability value of 0.43. However, with only a few stars it is difficult to statistically confirm this trend.

\begin{figure}[!ht]
		\centering
		\resizebox{\hsize}{!}{\includegraphics{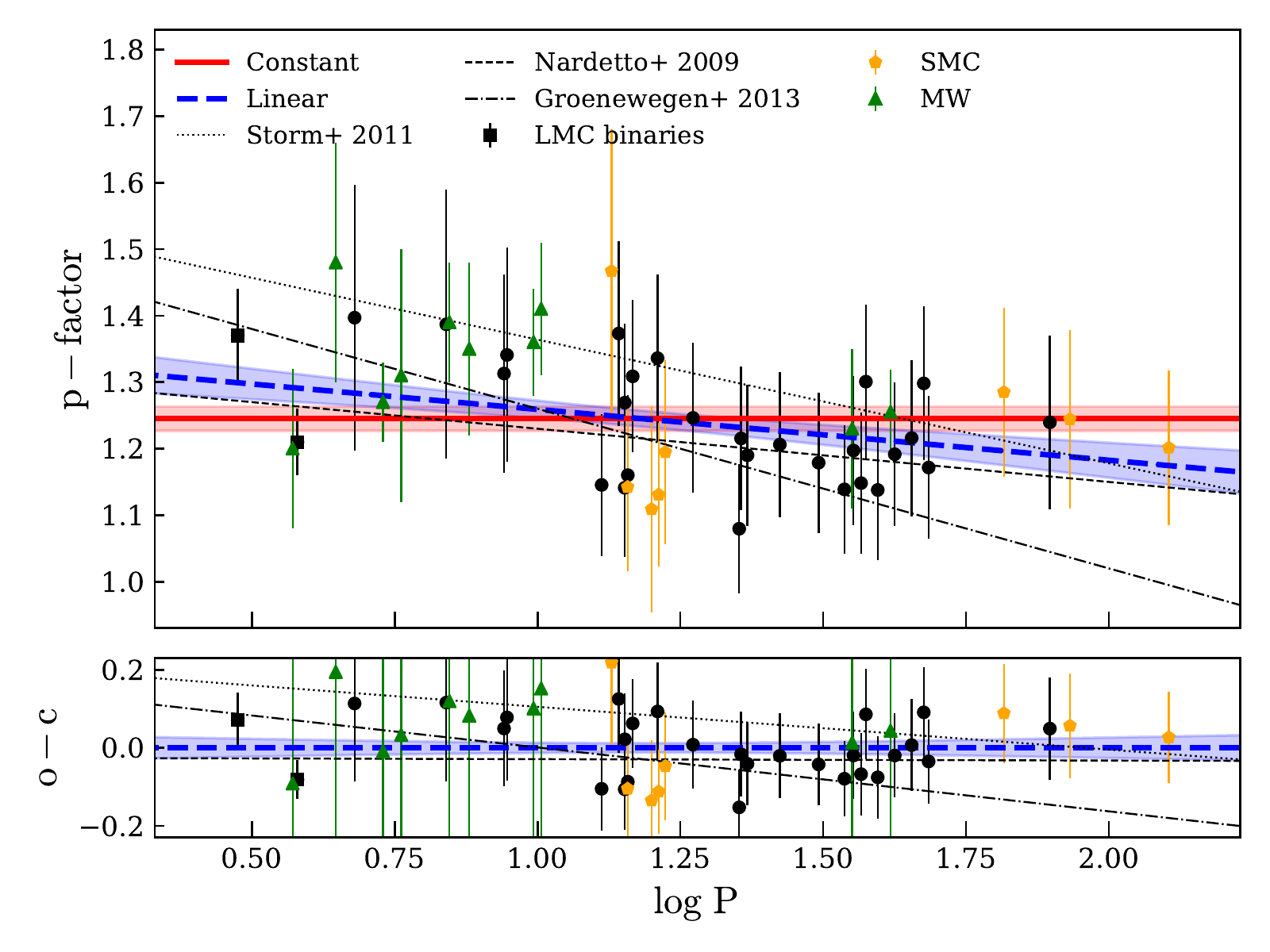}}
		\caption{Same as Fig.~\ref{figure_pfactor}, but including the all Cepheids, i.e. from the Milky way (green), the LMC (black), and the SMC (orange).}
		\label{figure_pfactor_all}
	\end{figure}

	\paragraph{- LMC+SMC+MW}: We now combine the Magellanic Cloud Cepheids with the Galactic ones. Galactic $p$-factors were taken from \citet[][1 Cepheid]{Kervella_2017_04_0}, \citet[][8 Cepheids, we rejected the known binary \object{FF~Aql}]{Breitfelder_2016_03_0} and \citet[][1 Cepheid]{Merand_2015_12_0}. They were all derived from a SPIPS analysis. The plot is presented in Fig.~\ref{figure_pfactor_all}. The result of the linear model gives
	\begin{displaymath}
		p = \slopeall_{\pm\eslopeall} (\log P - \logPOpPall) + \zeroall_{\pm\ezeroall},\ \sigma = 0.09,
	\end{displaymath}
	with a $\chi_r^2 = 0.61$. The fit of a constant value gives $p = \csteall \pm \ecsteall$ with $\chi_r^2 = 0.69$ (we again took the standard error). The $F$-test gives a probability of 0.02, and therefore strengthens the linear model, although our derived constant value is at $1\sigma$ with Kervella's $p$-factor. Our LMC+SMC+MW relation is parallel to our LMC relation and Nardetto's, but still within $1\sigma$. 
	
	A larger sample would certainly improve this analysis, and would also allow us to study possible metallicity effects on the $p$-factor. For now, the Galactic and SMC samples are too small to be able to investigate it properly. A first quick analysis, however, shows no strong evidence of a metallicity-dependent $p$-factor. The LMC mean $p$-factor is only at $1\sigma$ to the SMC and Kervella values, while the whole sample MW+LMC+SMC is at $1\sigma$ . We therefore agree with the conclusion of \citet{Groenewegen_2013_02_0} that there is no significant metallicity dependence. The next Gaia parallaxes will provide accurate parallaxes for hundreds of MW Cepheids, and will allow us to accurately calibrate the $P-p$ relation.
	
	
	\subsection{Period-radius relation}
	
	SPIPS also provides additional stellar parameters such as the mean effective temperature, luminosity, and linear radius. They are listed in Table~\ref{table__stellarParameter}. Linear radii were derived from the estimated angular diameters and the average distance of the Magellanic Clouds (again, a $\sim 8$\,\% error has to be added to take into account the average widths).
	
	From our multi-band surface brightness (SB) method,  we derived a new period-radius (P-R) relation for Cepheids, which is particularly interesting in order to constrain Cepheid models.  We added two others radii values derived from two LMC Cepheids in an eclipsing binary system \citep{Gieren_2015_12_0,Pilecki_2013_12_0}. We also extended the sample with radii of Galactic Cepheids determined with the same SPIPS method \citep{Breitfelder_2016_03_0}. All Cepheids are classical Cepheids, and therefore provide a uniform sample.
	
\begin{figure}[!ht]
		\centering
		\resizebox{\hsize}{!}{\includegraphics{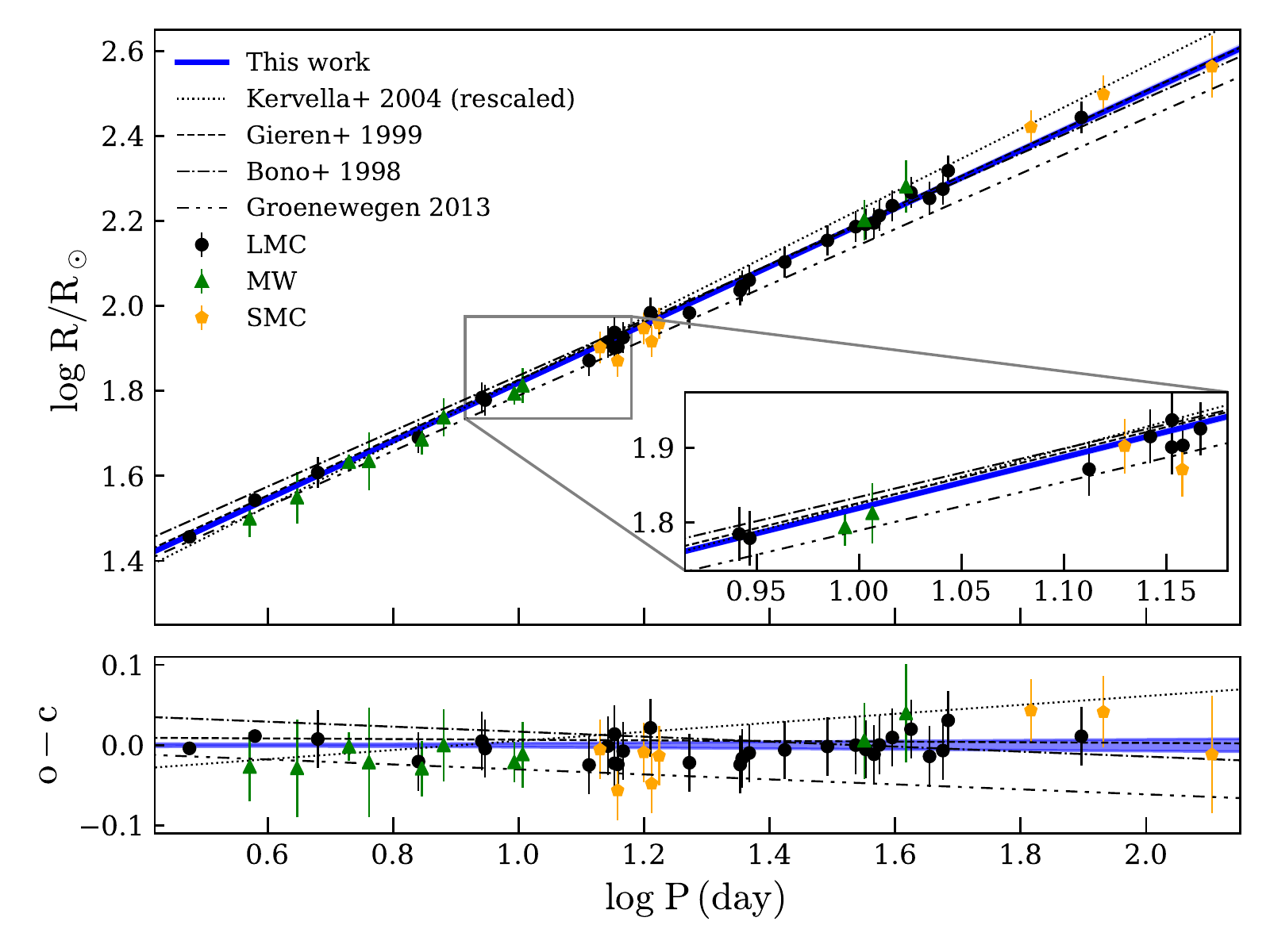}}
		\caption{Period-radius relation derived from our SPIPS analysis. The shaded blue and red areas denote the $1\sigma$ confidence interval.}
		\label{figure_PR_relation}
	\end{figure}

	The relation is plotted on Fig.~\ref{figure_PR_relation}. We performed a linear fit following our formalism of Sect.~\ref{subsection__linear_fit_and_uncertainties}, and derived a new $P-R$ relation:
	\begin{displaymath}
		\log R = \slopePR_{\pm\eslopePR} (\log P - \logPOPR) + \zeroPR_{\pm\ezeroPR},\ \sigma = 0.020.
	\end{displaymath}
	
	As shown in Fig.~\ref{figure_PR_relation}, we have a very good agreement with the empirical relation $0.680_{\pm0.017} \log P + 1.146_{\pm0.025}$ ($\sigma = 0.045$) of \citet[][]{Gieren_1999_02_0}, which was derived from a sample of 28 Galactic, 10 LMC, and 6 SMC Cepheids, and used a $(V-R)$ SB relation to derive angular diameters. Although they made used of a $P-p$ relation to estimate their linear radii, the level of consistency is very good, $\Delta a = +0.2\sigma_a$ and $\Delta b = -0.4\sigma_b$ with their error bars. The agreement with the theoretical relation $0.653_{\pm0.006} \log P + 1.183_{\pm0.009}$ of \citet{Bono_1998_04_0} is quite close, and looks almost parallel in our period range. The zero point of \citet[][$0.651_{\pm0.012} \log P + 1.136_{\pm0.014}, \sigma = 0.055$]{Groenewegen_2013_02_0}, who revisited the BW method using $(V-K)$ SB relation for 120 Galactic, 42 LMC and 6 SMC Cepheids, is in perfect agreement with our value, but the slope is at $2.6\sigma$. We note that the relation of \citet{Groenewegen_2013_02_0} is more dispersed than Gieren's relation, with $\sigma = 0.055$. The empirical relation of \citet{Kervella_2004_08_0} does not really match and is steeper. They applied the interferometric BW method, i.e. by combining direct measurements of angular diameters from interferometry with radial velocities. However, they assumed a constant $p = 1.36$ for their sample of nine Cepheids, which mostly explains the disagreement. By rescaling their relation with ours, we found a better agreement. The MW data used by \citet{Breitfelder_2016_03_0} include interferometric angular diameters \citep{Lane_2002_07_0,Kervella_2004_03_0,Gallenne_2012_03_0,Davis_2009_04_0} in addition to radial velocities, photometry, and trigonometric parallaxes measured with the Hubble Space Telescope (HST) Fine Guidance Sensor \citep{Benedict_2007_04_0}, and therefore lead to more robust and consistent estimates of the linear radii.
	
	\begin{sidewaystable*}[!h]
		\centering
		\caption{Fitted intrinsic parameters of the Magellanic Cloud Cepheids.}
		\begin{tabular}{ccccccccc}
			\hline\hline
			Star	& $P$    & MJD$_0$ & $\overline{\theta_\mathrm{LD}}$  &  $R$\tablefootmark{a}  & $\log \overline{T_\mathrm{eff}}$  &  $\log \overline{L/L_\odot}$ & $3.6\mu$m excess &  $4.5\mu$m excess  \\
			& (day)  & (day)        & ($\mu$as)   								    	   &  $R_\odot$ &  ($K$)	&  &   (mag) & (mag)    	\\
			\hline
			\multicolumn{7}{c}{LMC} \\
			HV6093  &  4.7850$\pm$0.0001 & 52175.93$\pm$0.07  & 7.5$\pm$0.2 & 40.3$\pm$1.0 &  3.76$\pm$0.02 & 3.22$\pm$0.09  &  --  &  -- \\ 
			HV2405  &  6.9238$\pm$0.0001 & 52163.10$\pm$0.13  & 9.1$\pm$0.2 & 48.9$\pm$1.1 &  3.75$\pm$0.01 & 3.33$\pm$0.05  &  $0.07\pm0.05$  &  $0.11\pm0.05$ \\ 
			HV12452  &  8.7386$\pm$0.0001 & 52164.52$\pm$0.08  & 11.3$\pm$0.3 & 60.7$\pm$1.4 &  3.76$\pm$0.02 & 3.57$\pm$0.07  &  $0.06\pm0.04$  &  $0.06\pm0.03$  \\ 
			HV12717  &  8.8432$\pm$0.0001 & 52158.71$\pm$2.47  & 11.2$\pm$0.3 & 60.2$\pm$1.5 &  3.76$\pm$0.02 & 3.54$\pm$0.08  &  $0.04\pm0.04$  &  $0.09\pm0.03$ \\ 
			HV2527  &  12.9491$\pm$0.0002 & 52163.87$\pm$0.20  & 13.8$\pm$0.3 & 74.2$\pm$1.7 &  3.73$\pm$0.03 & 3.63$\pm$0.10  &  $0.08\pm0.03$  &  $0.10\pm0.03$ \\ 
			HV2538  &  13.8726$\pm$0.0004 & 52166.39$\pm$0.63  & 15.3$\pm$0.4 & 82.2$\pm$2.1 &  3.73$\pm$0.02 & 3.71$\pm$0.09  &  $0.07\pm0.02$  &  $0.07\pm0.04$ \\ 
			HV5655  &  14.2118$\pm$0.0002 & 52161.68$\pm$0.06  & 14.8$\pm$0.3 & 79.5$\pm$1.8 &  3.72$\pm$0.01 & 3.65$\pm$0.06  &  $0.10\pm0.03$  &  $0.11\pm0.02$ \\ 
			HV1006  &  14.2159$\pm$0.0002 & 52173.64$\pm$0.14  & 16.1$\pm$0.4 & 86.5$\pm$2.0 &  3.73$\pm$0.02 & 3.74$\pm$0.08  &  $0.05\pm0.02$  &  $0.07\pm0.02$ \\ 
			HV12505  &  14.3899$\pm$0.0017 & 52160.85$\pm$0.16  & 14.9$\pm$0.4 & 80.1$\pm$1.9 &  3.72$\pm$0.02 & 3.63$\pm$0.08 &  $0.09\pm0.04$  &  $0.11\pm0.05$ \\ 
			HV2282  &  14.6767$\pm$0.0002 & 52157.29$\pm$0.08  & 15.7$\pm$0.4 & 84.4$\pm$1.9 &  3.74$\pm$0.01 & 3.75$\pm$0.05  &  $0.03\pm0.02$  &  $0.05\pm0.03$  \\ 
			HV2549  &  16.2220$\pm$0.0002 & 52158.54$\pm$0.29  & 18.0$\pm$0.4 & 96.7$\pm$2.3 &  3.74$\pm$0.02 & 3.89$\pm$0.10  &  $0.05\pm0.02$  &  $0.06\pm0.02$ \\ 
			HV1005  &  18.7144$\pm$0.0004 & 52179.27$\pm$0.10  & 17.9$\pm$0.4 & 96.2$\pm$2.2 &  3.73$\pm$0.01 & 3.83$\pm$0.06  &  $0.06\pm0.02$  &  $0.12\pm0.03$ \\ 
			U1  &  22.5450$\pm$0.0005 & 52146.16$\pm$0.14  & 20.2$\pm$0.5 & 108.6$\pm$2.6 &  3.72$\pm$0.01 & 3.93$\pm$0.06  &   $0.08\pm0.02$  &  $0.07\pm0.03$  \\ 	
			HV876  &  22.7156$\pm$0.0005 & 52145.62$\pm$0.21  & 20.7$\pm$0.5 & 111.2$\pm$2.7 &  3.74$\pm$0.03 & 4.01$\pm$0.13 &  $0.08\pm0.02$  &  $0.07\pm0.03$ \\
			HV878  &  23.3059$\pm$0.0004 & 51910.15$\pm$0.11  & 21.4$\pm$0.5 & 115.0$\pm$2.7 &  3.73$\pm$0.01 & 3.99$\pm$0.06  &  $0.06\pm0.02$  &  $0.06\pm0.02$  \\ 
			HV1023  &  26.5562$\pm$0.0007 & 51967.64$\pm$0.17  & 23.6$\pm$0.6 & 126.8$\pm$3.0 &  3.71$\pm$0.02 & 4.01$\pm$0.089  &  $0.04\pm0.02$  &  $0.03\pm0.02$ \\ 
			HV899  &  31.0518$\pm$0.0005 & 50719.62$\pm$0.14  & 26.5$\pm$0.6 & 142.4$\pm$3.3 &  3.70$\pm$0.01 & 4.06$\pm$0.04   &  $0.06\pm0.01$  &  $0.07\pm0.03$ \\ 
			HV873  &  34.4519$\pm$0.0006 & 51964.95$\pm$0.24  & 28.6$\pm$0.7 & 153.7$\pm$3.5 &  3.74$\pm$0.01 & 4.31$\pm$0.05  &  $0.06\pm0.01$  &  $0.05\pm0.03$ \\ 
			HV881  &  35.7366$\pm$0.0014 & 51865.24$\pm$0.25  & 29.0$\pm$0.7 & 155.8$\pm$3.7 &  3.71$\pm$0.01 & 4.18$\pm$0.06  &  $0.03\pm0.01$  &  $0.03\pm0.03$ \\ 
			HV879  &  36.8444$\pm$0.0012 & 51971.11$\pm$0.51  & 29.2$\pm$0.8 & 156.9$\pm$4.1 &  3.70$\pm$0.02 & 4.14$\pm$0.07  &  $0.07\pm0.02$  &  $0.06\pm0.03$ \\ 
			HV909  &  37.5658$\pm$0.0031 & 50010.95$\pm$0.27  & 30.4$\pm$0.7 & 163.4$\pm$3.8 &  3.72$\pm$0.01 & 4.27$\pm$0.04  &  $0.03\pm0.04$  &  $0.07\pm0.04$ \\ 
			HV2257  &  39.4039$\pm$0.0007 & 51923.52$\pm$0.30  & 32.0$\pm$0.8 & 172.0$\pm$4.0 &  3.70$\pm$0.01 & 4.23$\pm$0.05  &  $0.05\pm0.02$  &  $0.04\pm0.03$ \\ 
			HV2338  &  42.2002$\pm$0.0036 & 49988.16$\pm$0.34  & 34.4$\pm$0.8 & 184.9$\pm$4.4 &  3.70$\pm$0.01 & 4.30$\pm$0.04  &  $0.04\pm0.03$  &  $0.02\pm0.08$ \\ 
			HV877  &  45.1742$\pm$0.0033 & 51979.64$\pm$2.27  & 33.3$\pm$1.3 & 179.0$\pm$6.7 &  3.68$\pm$0.02 & 4.18$\pm$0.07  &  $0.05\pm0.01$  &  $0.02\pm0.04$ \\ 
			HV900  &  47.4361$\pm$0.0010 & 51989.19$\pm$0.60  & 35.0$\pm$0.9 & 188.1$\pm$4.8 &  3.70$\pm$0.03 & 4.30$\pm$0.14  &  $0.07\pm0.04$  &  $0.09\pm0.01$ \\ 
			HV2369  &  48.3688$\pm$0.0024 & 52012.57$\pm$0.40  & 38.7$\pm$0.9 & 208.0$\pm$4.9 &  3.71$\pm$0.01 & 4.44$\pm$0.05  &  $0.06\pm0.03$  &  $0.04\pm0.01$  \\ 
			HV2827  &  78.8533$\pm$0.0346 & 52088.82$\pm$6.51  & 51.7$\pm$1.4 & 277.8$\pm$7.3 &  3.69$\pm$0.01 & 4.60$\pm$0.05  &  $0.02\pm0.06$  &  $-0.03\pm0.11$  \\ 
			\hline
			\multicolumn{7}{c}{SMC} \\
			HV1345  &  13.4782$\pm$0.0002 & 50622.16$\pm$0.33  & 12.0$\pm$0.4 & 79.9$\pm$2.5 &  3.72$\pm$0.02  &  3.64$\pm$0.07  &  $0.01\pm0.05$  &  $0.07\pm0.05$ \\ 
			HV1335  &  14.3808$\pm$0.0002 & 50624.71$\pm$0.47  & 11.1$\pm$0.3 & 74.3$\pm$2.3 &  3.76$\pm$0.02  &  3.73$\pm$0.10   &  $0.04\pm0.03$  &  $0.06\pm0.05$ \\ 
			HV1328  &  15.8369$\pm$0.0007 & 52078.24$\pm$0.77  & 13.2$\pm$0.4 & 88.5$\pm$2.8 &  3.76$\pm$0.02  &  3.90$\pm$0.10  &  $0.03\pm0.04$  &  $0.05\pm0.03$ \\ 
			HV1333  &  16.2952$\pm$0.0001 & 52085.07$\pm$0.20  & 12.4$\pm$0.4 & 82.5$\pm$2.6 &  3.73$\pm$0.01  &  3.71$\pm$0.06  &  $0.06\pm0.03$  &  $0.09\pm0.04$ \\ 
			HV822  &  16.7416$\pm$0.0001 & 50616.44$\pm$0.15  & 13.6$\pm$0.4 & 91.0$\pm$2.9 &  3.72$\pm$0.02  &  3.75$\pm$0.08  &  --  &  -- \\ 
			HV824  &  65.5188$\pm$0.0635 &  52051.00$\pm$1.39  & 39.5$\pm$1.7 & 263.5$\pm$11.1 &  3.71$\pm$0.01  &  4.65$\pm$0.06  &  $0.10\pm0.15$  &  $0.14\pm0.12$ \\
			HV829  &  85.4960$\pm$0.2340 &  52046.81$\pm$4.52  & 47.1$\pm$3.1 & 314.8$\pm$20.5 &  3.73$\pm$0.01  &  4.86$\pm$0.08  &  $0.01\pm0.22$  &  $0.05\pm0.19$ \\ 
			HV821  &  127.3405$\pm$0.0276 &  44131.5$\pm$27.2  & 54.8$\pm$8.1 & 365.8$\pm$54.2 &  3.69$\pm$0.01  &  4.84$\pm$0.14  &  --  &  -- \\ 
			\hline& 
		\end{tabular}
		\tablefoot{a) The quoted errors are only statistical, a 8\,\% error has to be added to take into account the average width of the LMC, and 7.9\,\% for the SMC.}
		\label{table__stellarParameter}
	\end{sidewaystable*}
	
	\subsection{Mid-infrared excess}
	
	A significant fraction of Galactic classical Cepheids exhibits an IR excess that is probably caused by a circumstellar envelope (CSE). The discovery of the first CSE around the Cepheid $\ell$~Car made use of near- and mid-IR interferometric observations \citep{Kervella_2006_03_0}. Similar detections were then reported for other Galactic Cepheids in the IR \citep{Merand_2007_08_0,Barmby_2011_11_0,Gallenne_2011_11_0,Gallenne_2013_10_0} and visible \citep{Nardetto_2017_01_0}, leading to the hypothesis that maybe all Cepheids are surrounded by a CSE. The origin of these envelopes is still unknown. They might be related to past or ongoing stellar mass loss, and might be used to trace the Cepheid evolution history. 
	
	The main issue regarding the presence of these CSEs is that they might induce a bias to distance determinations made with the Baade-Wesselink method, and bias the calibration of the IR period–luminosity relation. Our previous works \citep{Merand_2006_07_0,Gallenne_2011_11_0,Gallenne_2013_10_0} showed that these CSEs have an angular size of a few stellar radii and a flux contribution to the photosphere ranging from 2\,\% to 30\,\%. While in the near-IR the CSE emission might be negligible compared with the stellar continuum, this is not the case in the mid- and far-IR, where the CSE emission dominates. \citet{Merand_2007_08_0} and \citet{Gallenne_2011_11_0,Gallenne_2013_10_0} pointed out a possible correlation between the pulsation period and the CSE brightness in the near- and mid-IR. It seems that long-period Cepheids have brighter CSE than shorter period Cepheids. This could indicate a mass-loss mechanism linked to the stellar pulsation, and therefore that heavier stars experience higher mass-loss rate. This also seems to be consistent with the theoretical work of \citet{Neilson_2008_09_0} who found a correlation between the mass-loss rate and the pulsation period.
	
\begin{figure}[!ht]
		\centering
		\resizebox{\hsize}{!}{\includegraphics{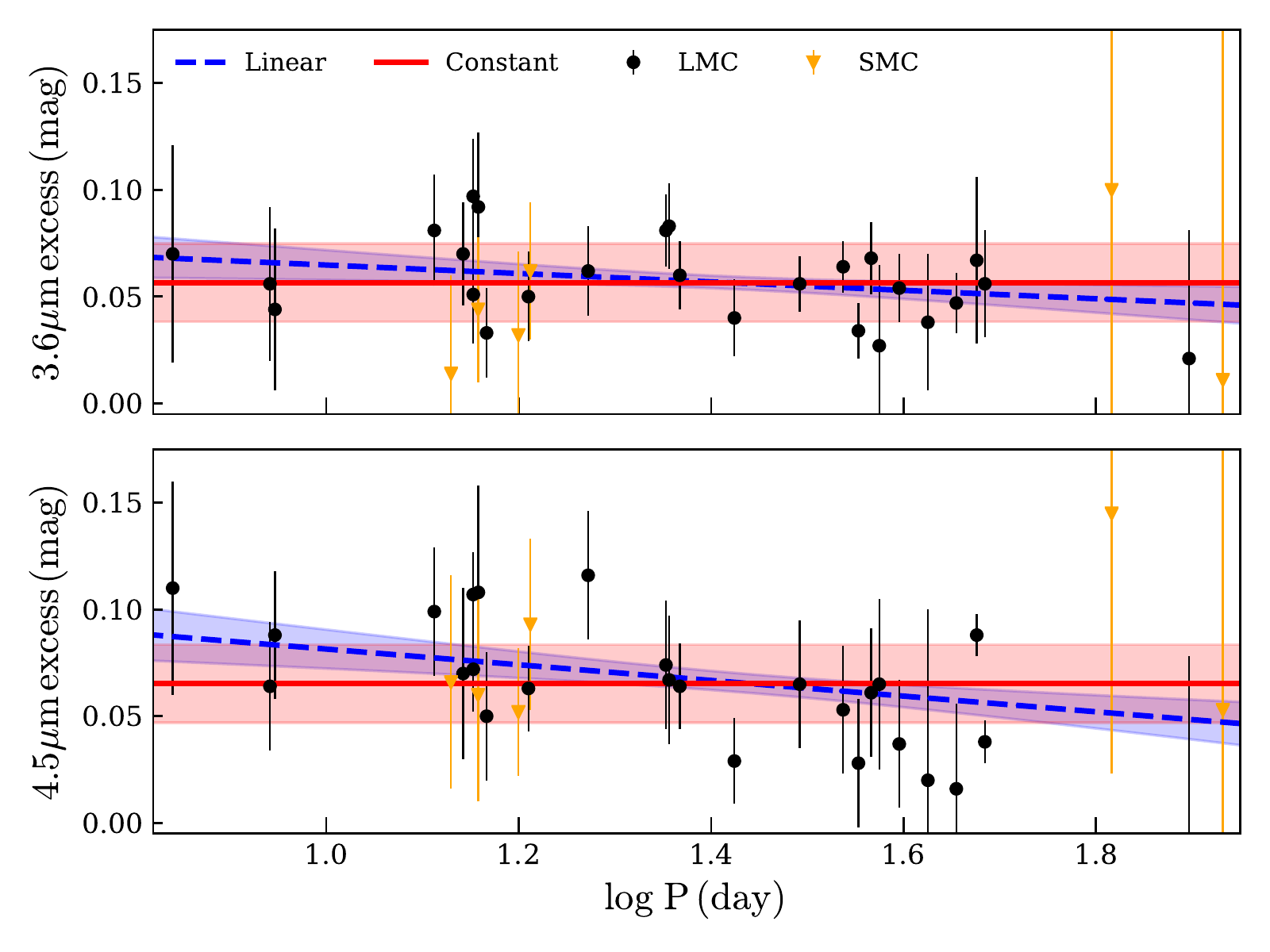}}
		\caption{Infrared excess with respect to pulsation period at $3.6\mu$m and $4.5\mu$m.}
		\label{figure_IRexcess}
	\end{figure}

	As discussed in Sect.~\ref{section__spips_analysis}, our SPIPS fits from $B$ to $K$ band provide estimates of IR excess at $3.6\mu$m and $4.5\mu$m for Cepheids having Spitzer light curves \citep{Scowcroft_2011_12_0}. The derived excess magnitudes are listed in Table~\ref{table__stellarParameter}. Examples of our final fit are shown in Figs.~\ref{figure_HV12452}, \ref{figure_HV900}, and \ref{figure_HV1333}. We clearly see an offset above the photospheric level (in grey), and the stars appear brighter due to the presence of IR excess. In Fig.~\ref{figure_IRexcess} we plotted those IR excess with respect to the pulsation period, and we note that all stars exhibit an excess. However, the trend seems constant with the period at these wavelengths, which differs from our previous studies at $2.2\,\mu$m and $8.6\,\mu$m for Galactic Cepheids.
	
	We performed a linear and constant fit for both wavelengths combining all Magellanic Cloud Cepheids. From an F-test, we found that a constant model is more statistically significant. We estimated an average excess of $\Delta m_{3.6} = 0.057 \pm 0.010$\,mag and $\Delta m_{4.5} = 0.065 \pm 0.010$\,mag (again, we took the standard deviation). These excesses are particularly significant and might be the signature of circumstellar dust. 
	
	These results are relevant for the mid-IR calibration of the Cepheid distance scale \citep[see e.g.][]{Freedman_2012_10_0}. Some authors find that there are several advantages to using a mid-infrared $P-L$ relation. The effect of the dust extinction is reduced, the intrinsic dispersion is smaller than in the $V$ band, and the metallicity effect is supposed to be minimal. However, going to longer wavelengths brings new systematics that impact the distance scale, and which are usually not taken into account. Our measured IR excesses convert to a systematic error of 2.6\,\% and 3\,\% on the distance at $3.6\,\mu$m and $4.5\,\mu$m, respectively. Unfortunately, recent papers \citep{Scowcroft_2016_01_0,Freedman_2012_10_0,Monson_2012_11_0} do not take into account this additional uncertainty. On the other hand, their relation is calibrated for a Cepheid plus a CSE, which is still valid if we assume that all Cepheids have an IR excess emission. The presence of IR excess probably also contributes to the dispersion of this relation.
		
	\section{Conclusion}
	\label{section__conclusion}
	
	We performed SPIPS modelling of Magellanic Cloud Cepheids using its average distances estimated in \citet{Pietrzynski_2013_03_0} and \citet{Graczyk_2014_01_0} to empirically derive the projection factor of each star. Combining with MW results from \citet{Kervella_2017_04_0}, \citet{Breitfelder_2016_03_0} and \citet{Merand_2015_12_0}, we confirm a linear relation of the $p$-factor with the pulsation period, consistent with the last results of \citet{Nardetto_2009_05_0}. This relation shows a decreasing $p$-factor with the period. We note, however, that the SMC sample contains only eight Cepheids with low quality data; new accurate data will probably improve the calibration. The MW sample will also be increased thanks to the upcoming Gaia parallaxes; thanks to this new data we will be able to apply our SPIPS analysis to a very large sample of MW Cepheids.
	
	A new calibration of the period-radius relation for Galactic and Magellanic Cloud Cepheids has been derived, based on ATLAS9 models. This relation is consistent with previous works, but with a smaller intrinsic dispersion of $\pm 0.02$, i.e. 4.5\,\% in radius. Our work also suggest that there is one universal $P-R$ relation for the MW and Magellanic Cloud Cepheids. 
	
	Infrared excesses at $3.6\mu$m and $4.5\mu$m have been detected, which might be the signature of the presence of circumstellar dust. There is no linear trend, which differs from our previous work at $2.2\,\mu$m and $8.6\,\mu$m. A systematic offset of $\sim 0.06$\,mag is estimated. This has a potential impact on the Hubble constant calibrated at those wavelengths, where circumstellar envelopes of Cepheids are not taken into account. The James Webb Space Telescope (JWST) will operate in the mid-IR and is supposed to measure $H_0$ to 1\,\% from Cepheids observation, but the presence of CSE is likely to bias the distance scale, making the mid-IR $P-L$ relations not the ideal tool to accurately measure the Hubble constant and other derived cosmological parameters. 

	Gaia will soon provide accurate parallaxes for hundreds of Cepheids and will greatly improve the calibration of the zero point of the P-L relations, but the compensation of the interstellar reddening is likely to significantly limit the accuracy of this calibration. Reddening-independent interferometric angular diameters combined with multi-band photometry in our SPIPS modelling will better constrain the reddening. Interferometric measurements are therefore a particularly valuable observable for the determination of the Cepheid luminosities. In the near future, Gaia will enable us to calibrate accurately the P-p relation, therefore improving the usability of the Baade-Wesselink technique to determine the distances of distant Cepheids.

	%--------------------ACKNOWLEDGEMENTS--------------------
	
	\begin{acknowledgements}
		
		The authors acknowledge the support of the French Agence Nationale de la Recherche (ANR), under grant ANR-15-CE31-0012-01 (project Unlock-Cepheids). P.K., A.G., and W.G. acknowledge support of the French-Chilean exchange programme ECOS-Sud/CONICYT (C13U01). W.G. and G.P. gratefully acknowledge financial support for this work from the BASAL Centro de Astrofisica y Tecnologias Afines (CATA) PFB-06/2007. W.G. also acknowledges financial support from the Millenium Institute of Astrophysics (MAS) of the Iniciativa Cientifica Milenio del Ministerio de Economia, Fomento y Turismo de Chile, project IC120009. We acknowledge financial support from the Programme National de Physique Stellaire (PNPS) of CNRS/INSU, France. The research leading to these results has received funding from the European Research Council (ERC) under the European Union’s Horizon 2020 research and innovation programme (grant agreement No. 695099). This work made use of the SIMBAD and VIZIER astrophysical database from CDS, Strasbourg, France and the bibliographic information from the NASA Astrophysics Data System.
		
	\end{acknowledgements}
	
	\bibliographystyle{aa}   % if natbib is available
	\bibliography{/Users/agallenn/Sciences/Articles/bibliographie}
	
	\begin{appendix}
		\section{SPIPS model for 2 LMC and 2 SMC Cepheids}
		\label{appendix__1}
		The figures in this appendix show the result of the SPIPS modelling of four stars from our sample. In all plots, the model is represented using a grey curve. The number $i$ in the spline comb denotes the $i$th node. The Y-axis for the photometry (right panels) is listed in magnitude. The point on the right side of the photometric plots gives the error bar scale.
		\begin{figure*}[]
			\centering
			\resizebox{.9\hsize}{!}{\includegraphics{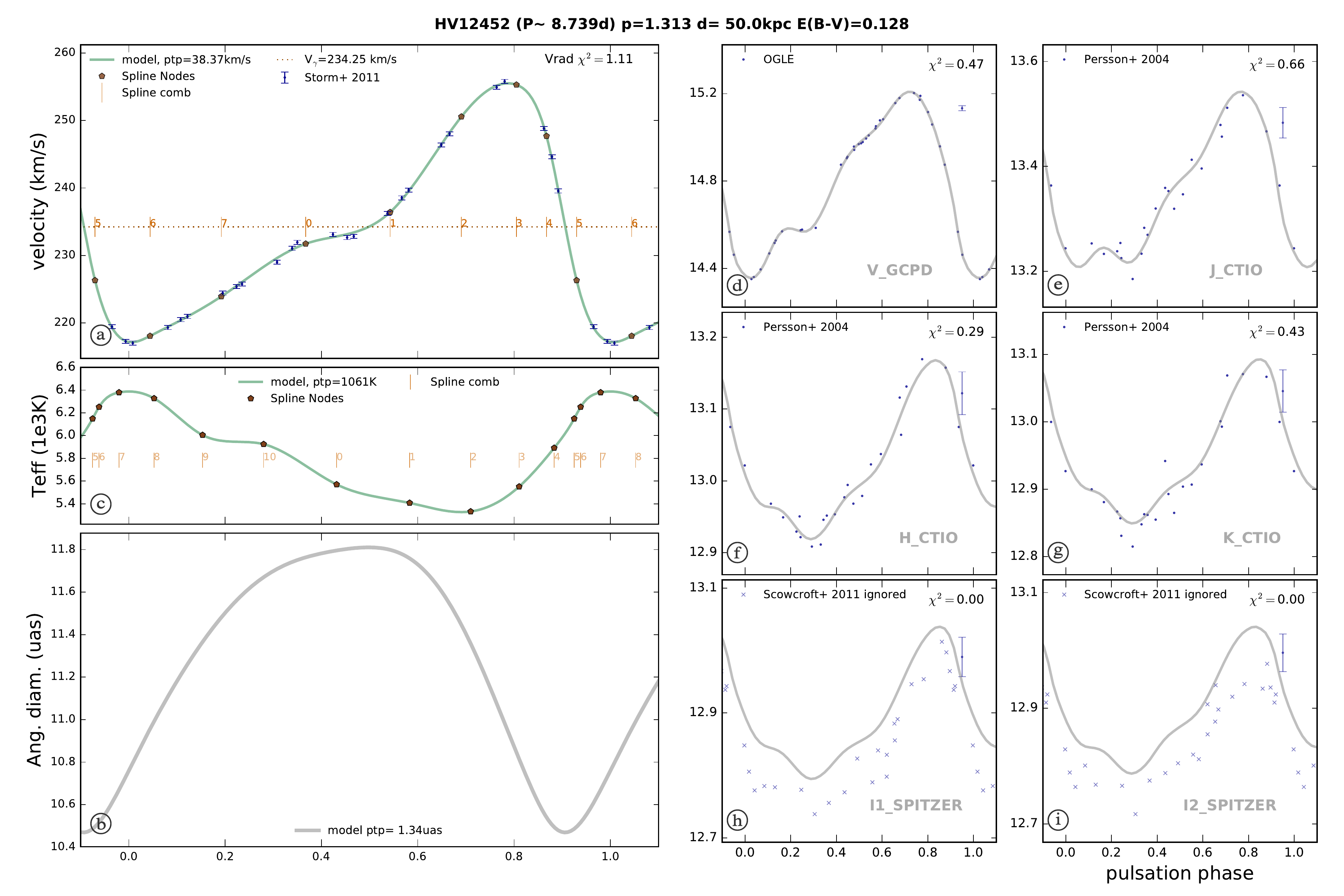}}
			\caption{SPIPS model for LMC Cepheid HV12452.}
			\label{figure_HV12452}
		\end{figure*}
		\begin{figure*}[]
			\centering
			\resizebox{.9\hsize}{!}{\includegraphics{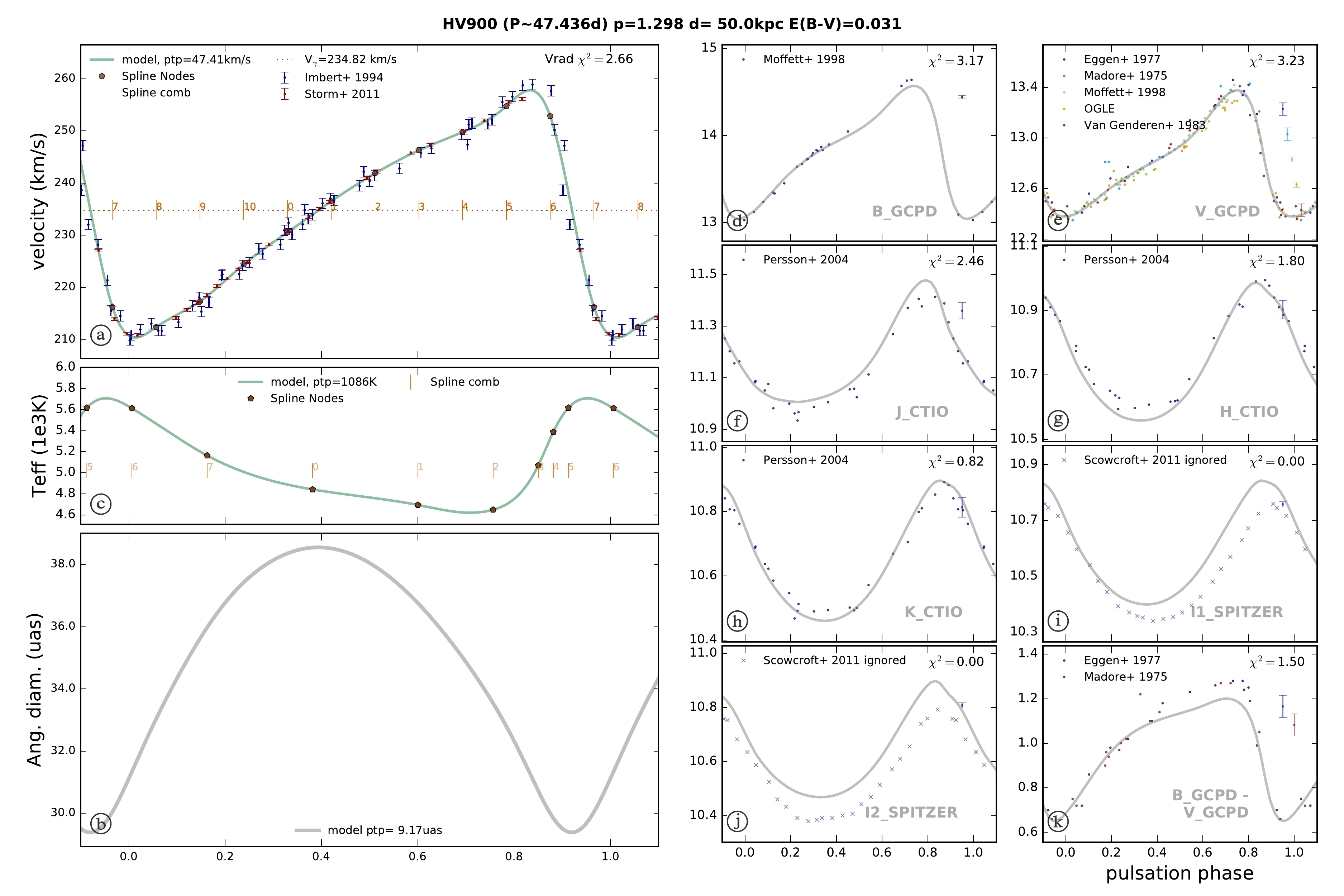}}
			\caption{SPIPS model for LMC Cepheid HV900.}
			\label{figure_HV900}
		\end{figure*}
		\begin{figure*}[]
			\centering
			\resizebox{.9\hsize}{!}{\includegraphics{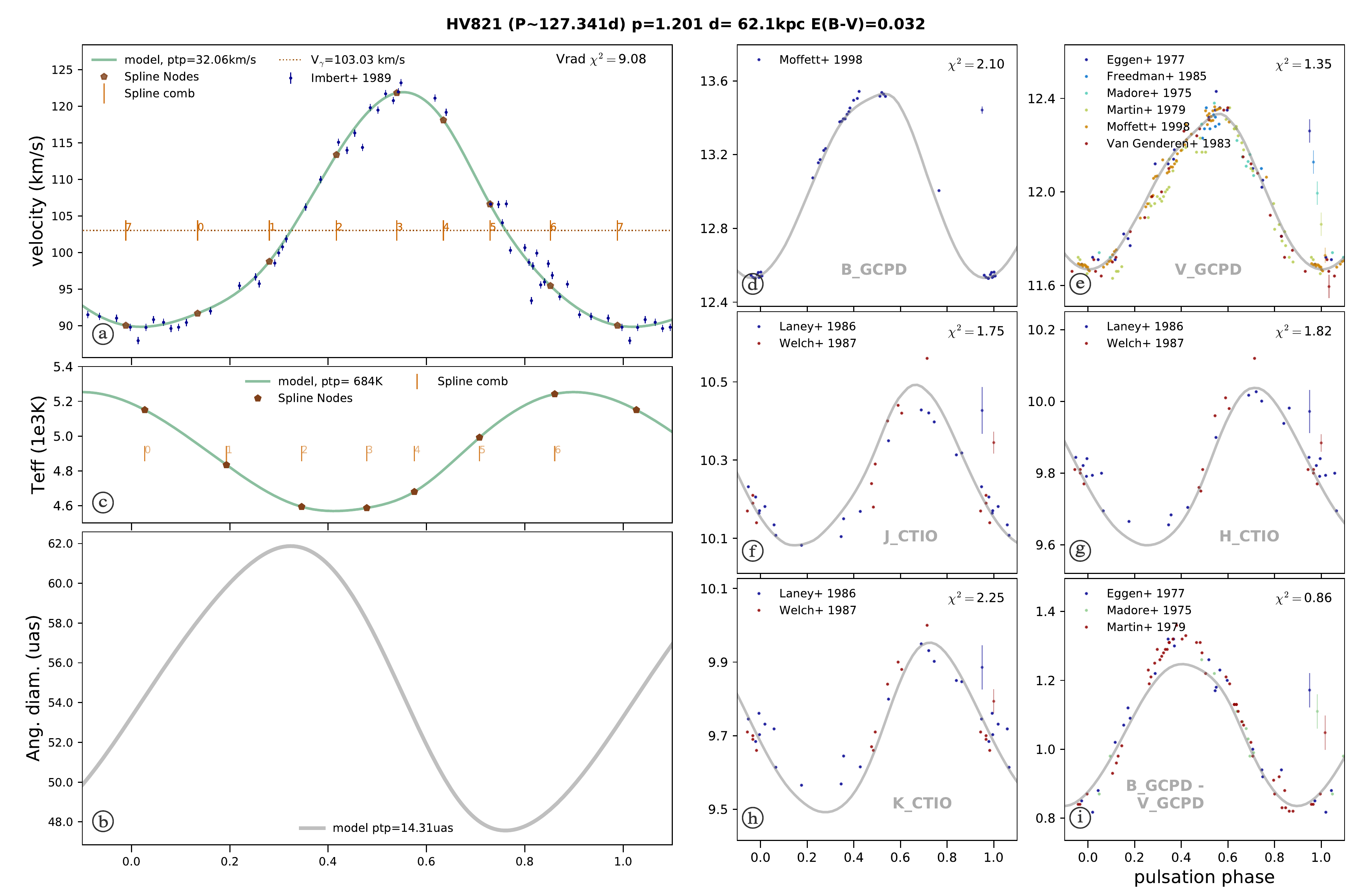}}
			\caption{SPIPS model for SMC Cepheid HV821.}
			\label{figure_HV821}
		\end{figure*}
		\begin{figure*}[]
			\centering
			\resizebox{.9\hsize}{!}{\includegraphics{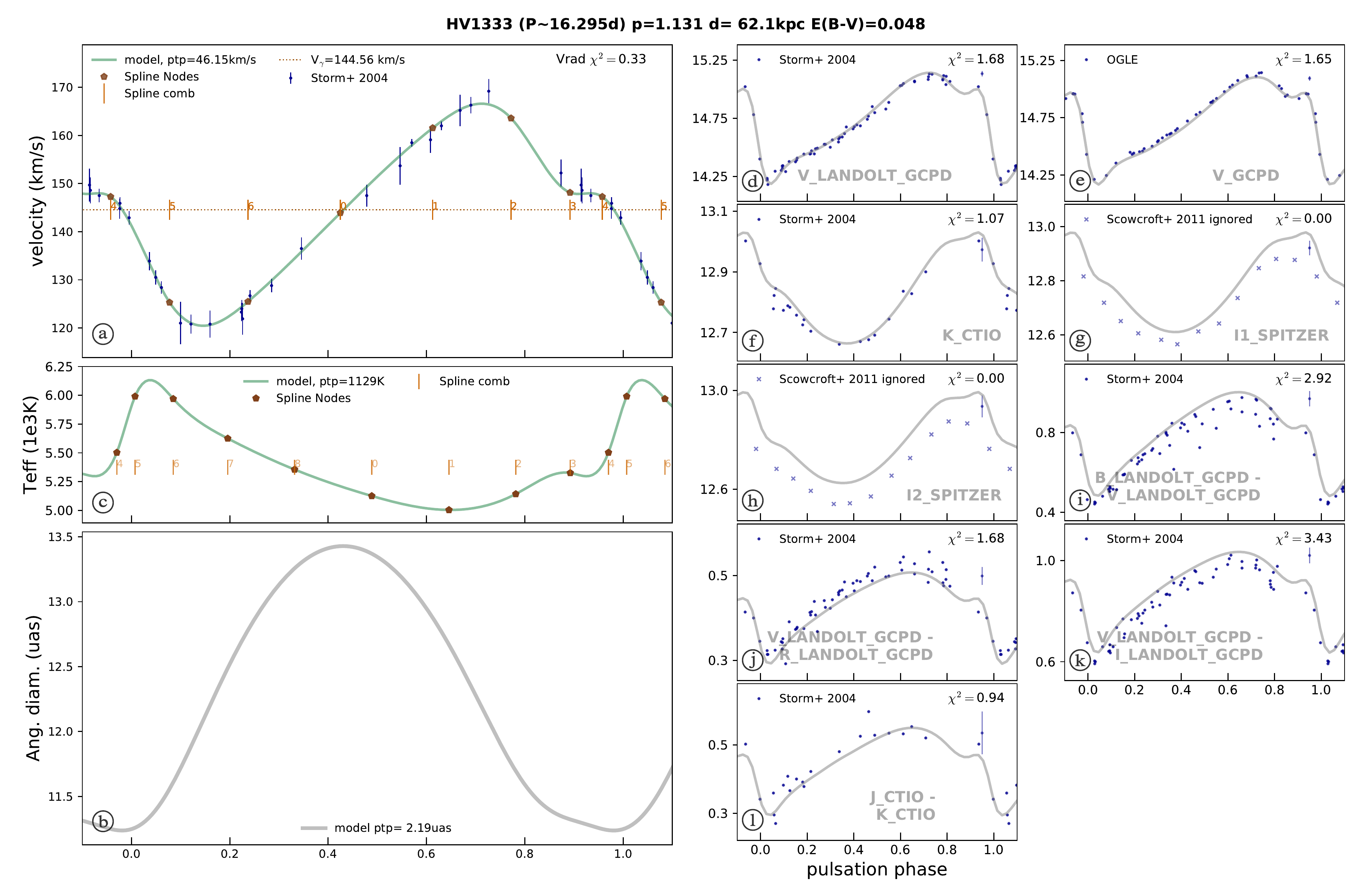}}
			\caption{SPIPS model for SMC Cepheid HV1333.}
			\label{figure_HV1333}
		\end{figure*}
	\end{appendix}
	
\end{document}